# A Diagnostics-First Composite Index for Macro-Financial Resilience to Socioeconomic Challenges: The Gondauri Index with Benchmarking and Scenario Evidence

**Davit Gondauri (corresponding author,** Dgondauri@gmail.com**),** Dr, Professor, Doctor of Business Administration, Business & Technology University, Georgia

**Type of manuscript:**
research paper

**Received** 12 of January, 2026

**Accepted** 15 of March, 2026

**Published** 31 of March, 2026

**Founder and Publisher:** Academic Research and Publishing UG (i.G.), Germany

**Abstract:** *In the face of socioeconomic challenges, this paper develops and empirically demonstrates the Gondauri Index (GI) as a reproducible, diagnostics-first composite framework for benchmarking macro-financial resilience across heterogeneous economies on a unified 0–100 scale. The GI is designed to address a central limitation of conventional surveillance dashboards: resilience is multi-dimensional and only partially substitutable, meaning that strength in one area cannot sustainably offset structural fragility in another. To operationalize this principle, GI integrates three interpretable pillars: Inequality Resilience Score (IRS) capturing the stability and macro-consistency of distributional dynamics; Liquidity & Systemic Resilience (LNSR) capturing the robustness of liquidity transmission and residual stress under rolling conditions; and Inflation Forecast Coherence (IFC) measuring whether inflation forecasting performance improves consistently when moving from benchmark specifications to a hybrid model with an FPAS + ζ augmentation. Cross-economy comparability is ensured through robust percentile normalization (p5–p95), a consistent annual country-year evaluation design, and explicit missing-data handling via component-level weight re-normalization. Empirically, the results provide a 2024 benchmark snapshot and dynamic evidence over 2005–2024 using 5-year rolling diagnostics and Δlog(GI) contribution decomposition, enabling transparent attribution of resilience changes to pillar-level drivers. A forward-looking extension constructs 2026–2030 scenario pathways (baseline/adverse/optimistic) and introduces a binding-pillar diagnostic that identifies the dominant constraint on resilience across horizons. Overall, the GI framework offers a scalable measurement tool for comparative resilience assessment, early-warning diagnostics, and evidence-based policy sequencing in both advanced and emerging economies facing socioeconomic challenges.*

**Cite as:** Gondauri, D. (2026). A Diagnostics-First Composite Index for Macro-Financial Resilience to Socioeconomic Challenges: The Gondauri Index with Benchmarking and Scenario Evidence. *SocioEconomic Challenges, 10*(1), 50–83. https://doi.org/10.61093/sec.10(1).50–83.2026.

# INTRODUCTION

Macroeconomic and financial resilience has become a central policy concern in an environment characterized by persistent inflation volatility, tightening global financial conditions, commodity-price shocks, and heightened geopolitical uncertainty. For small open economies, resilience is not merely a function of long-run growth potential; it is the capacity to absorb shocks while preserving macroeconomic stability, social cohesion, and the credibility of the nominal anchor. Episodes of global risk repricing, abrupt changes in energy and food prices, exchange-rate pass-through, and sudden stops in capital flows can translate quickly into distributional stress, liquidity shortages, and destabilizing inflation dynamics. In this setting, policymakers face socioeconomic challenges that are simultaneously macro-financial and social: vulnerabilities propagate across households, firms, and financial intermediaries, and the costs of adjustment are often borne unevenly across income groups. The result is a resilience problem that cannot be evaluated through any single indicator, because weaknesses tend to be nonlinear, mutually reinforcing, and time-varying.

Despite the salience of these socioeconomic challenges, the empirical toolkit used to compare resilience across heterogeneous economies remains fragmented. Conventional surveillance dashboards (tracking inflation, growth, public debt, unemployment, external balances, and inequality) are indispensable, yet they rarely produce a coherent and diagnostically interpretable summary measure that supports systematic benchmarking. At the other extreme, many composite indices provide ordinal rankings but often at the expense of interpretability, reproducibility, and a transparent mapping between changes in index values and underlying economic mechanisms. A ranking can signal "who is above whom," but it often fails to clarify why a country's resilience score changes, which domain deteriorated, and which policy transmission channels plausibly explain the shift. Moreover, composite frameworks frequently embed hidden choices about scaling, thresholds, weights, and treatment of outliers, which can make results difficult to replicate and to use operationally. These limitations become especially costly when the objective is early diagnostics: under socioeconomic challenges, policy must act before vulnerabilities become crises, so measurement frameworks must be both analytically rigorous and decision-relevant.

A second limitation of existing approaches is that resilience is intrinsically multidimensional and only partially substitutable. Strong performance in one domain cannot permanently compensate for structural fragility in another. For example, stable inflation and adequate headline growth can coexist with distributional stress that undermines social cohesion; similarly, a favorable debt ratio can coexist with liquidity fragilities that amplify shocks through the financial system. This partial substitutability is particularly relevant for small open economies, where adverse external conditions can expose latent weaknesses quickly and force pro-cyclical adjustments. Under socioeconomic challenges, such imbalances often appear as "quiet" vulnerabilities – rising inequality pressures, tightening liquidity transmission, or deteriorating inflation-forecast coherence – that are not immediately visible in headline indicators but materially shape the economy's shock-absorption capacity.

To address these gaps, this paper develops the Gondauri Index (GI) as a composite macro-financial resilience metric on a 0–100 scale, designed explicitly to support cross-country benchmarking and within-country diagnostics. The GI is constructed to be transparent and reproducible: it relies on clearly specified indicator transformations, robust normalization, and an architecture that allows decomposition into interpretable components rather than producing an opaque ranking. The guiding principle is a diagnostics-first framework that transforms a set of widely used macro-financial and distributional signals into a structured resilience assessment that can be tracked over time. By design, the GI emphasizes interpretability: a change in the overall index can be traced to a specific pillar and further to the underlying indicators and their standardized movements.

Methodologically, the GI implements multidimensional resilience through a three-pillar architecture: (i) inequality resilience, capturing the distributional stability and social-cohesion dimension; (ii) liquidity/systemic resilience, capturing the capacity of the financial system and liquidity transmission mechanisms to withstand stress without nonlinear amplification; and (iii) inflation-forecast coherence, capturing the credibility and informational consistency of the inflation process as reflected in forecasting performance and coherence metrics. These pillars are aggregated through a weighted geometric rule that penalizes imbalance: improvements in one pillar cannot fully offset deterioration in another, consistent with the economic logic of partial substitutability. This aggregation choice is not cosmetic; it directly encodes the idea that resilience failures typically arise from weak links that propagate across domains: an especially common feature of socioeconomic challenges, where macro-financial stress and distributional pressures interact.

The contribution of the paper is therefore twofold. First, it provides a structured index architecture that links distributional stability, liquidity transmission, and inflation-forecast coherence into a unified resilience score with transparent decomposition. Second, it operationalizes early diagnostics by prioritizing robustness, comparability, and interpretability: properties that are often traded off in existing composite measures. In practical terms, the GI is intended to support (a) cross-country benchmarking of macro-financial resilience under common scaling, (b) within-country monitoring of resilience trajectories across time, and (c) policy-relevant diagnostics that identify which pillar is driving changes in the resilience profile.

This study responds to socioeconomic challenges by proposing a reproducible, policy-oriented measurement framework that integrates key macro-financial mechanisms into a single diagnostics-first index while preserving transparency. By making the decomposition explicit and by penalizing structural imbalance, the GI is designed to keep the focus on actionable vulnerabilities rather than on ranking for its own sake. The remainder of the paper develops the conceptual logic of the index, specifies its construction and aggregation rules, and demonstrates how the framework can be used as an operational tool for benchmarking and early diagnostics of resilience to socioeconomic challenges.

## LITERATURE REVIEW

### Macro-Financial Resilience as a Multi-Dimensional Construct

Macroeconomic resilience has evolved from a descriptive concept into an operational object in central-bank surveillance, macroprudential monitoring, and fiscal-risk management. Contemporary interpretations emphasize the ability of an economy to absorb shocks while preserving core system functions, namely the stability of nominal anchors, the continuity of financial intermediation, and the maintenance of social cohesion (Briguglio et al., 2009; Duval & Vogel, 2008; Rose, 2004). In small open economies, resilience is increasingly seen as a joint outcome of stabilization buffers, structural flexibility, and institutional credibility (IMF, 2019; OECD, 2023).

Institutional quality is also relevant, since governance effectiveness, regulatory quality, and control of corruption shape a country's capacity to absorb and manage macro-financial shocks. This perspective is also consistent with the broader idea of antifragility, which distinguishes systems that merely resist shocks from those that adapt and potentially improve under disorder and stress (Taleb, 2012).

A central methodological challenge is that resilience is intrinsically multi-dimensional and only partially substitutable: strength in one domain cannot fully compensate for fragility in another. This motivates composite frameworks that move beyond indicator dashboards and provide diagnostic interpretation of vulnerability accumulation and shock transmission (Holló et al., 2012; Illing & Liu, 2006). In this setting, the Gondauri Index (GI) integrates inequality dynamics, liquidity/systemic transmission, and inflation forecast credibility into a single diagnostic composite metric, supporting both benchmarking and policy sequencing.

### Composite Indicators in Economics: From Rankings to Diagnostics

Composite indicators are widely used in economics, finance, governance, and development monitoring to summarize complex systems into interpretable metrics while preserving transparency about underlying drivers (Nardo et al., 2005; OECD, 2023). In macro-financial contexts, composite frameworks have become increasingly relevant for financial stability monitoring and early-warning analysis (IMF, 2021).

Because cross-country macro-financial datasets are often incomplete, rigorous treatment of missing observations is also a central methodological requirement in any composite framework intended for comparative use (Little & Rubin, 2019). Sensitivity analysis is likewise essential in composite-index design because it tests whether rankings and pillar contributions remain stable under alternative assumptions, parameter values, and scaling choices (Saltelli et al., 2005).

Methodological debates focus on aggregation rules, normalization under structural heterogeneity, and robustness to weighting assumptions. Linear aggregation is intuitive but highly compensatory, allowing severe weakness in one dimension to be masked by strength elsewhere. For resilience measurement, geometric aggregation is often preferred because it penalizes imbalance and aligns with the principle that minimum capacity across essential pillars is required for stability (Mazziotta & Pareto, 2013; Greco et al., 2018). Outlier-robust normalization (e.g., percentile-bounded scaling and winsorization) further improves comparability when distributions exhibit heavy tails and crisis outliers.

## Inequality Resilience and the Macro-Distributional Channel (IRS)

Inequality is increasingly treated as a macro-financial variable rather than a purely social outcome. A substantial literature links distributional instability to weaker long-run growth, reduced policy legitimacy, and higher probability of social-political shocks that propagate into financial fragility (Atkinson, 2015; Milanovic, 2016; Piketty, 2014). Importantly, inequality dynamics are shaped by inflation regimes, labor-market transitions, and growth volatility, which implies that inequality risk is not only a level phenomenon but also a stability and shock-sensitivity phenomenon (Blanchard & Rodrik, 2021; Solt, 2020). Recent evidence also shows that innovation-driven growth can coincide with greater concentration at the top of the income distribution, reinforcing the argument that inequality dynamics must be treated as structurally linked to modern economic transformation (Aghion et al., 2019). This distributional perspective is reinforced by research showing that inequality evolves through complex interactions between domestic structures and global integration, rather than through purely national drivers alone (Bourguignon, 2015).

Comparative evidence on poverty and inequality also confirms that distributional outcomes differ markedly across countries and over time, underscoring the need for frameworks that capture inequality dynamics rather than static levels alone (Ferreira & Ravallion, 2008). IMF evidence similarly suggests that rising inequality can weaken recovery processes and undermine macroeconomic resilience, especially when shocks are transmitted unevenly across households and labor-market groups (Furceri et al., 2018). Galbraith (2012) likewise argues that inequality and macroeconomic instability are deeply connected, with distributional imbalances often preceding wider systemic stress.

Inflation is frequently distributionally non-neutral, redistributing resources through nominal rigidities, heterogeneous balance sheets, and unequal access to hedging instruments (Coibion et al., 2017). Similarly, unemployment shocks disproportionately affect lower-income groups and can generate persistent inequality increases (OECD, 2015). These channels justify modelling inequality resilience as a component of macro-financial resilience rather than an external background variable. Within this stream, Gondauri (2024) operationalizes inequality dynamics through a Ricci-flow-inspired approach that interprets stability as smoothing and fragility as curvature-like shock amplification. This perspective motivates the GI's Inequality–Ricci Subindex (IRS) as a structured resilience pillar, focusing on the stability and regularity of distributional evolution.

At a deeper conceptual level, the Ricci-flow intuition used in the IRS pillar ultimately traces back to Perelman's entropy-based treatment of geometric evolution, which provides the mathematical inspiration for interpreting stability as smoothing and instability as curvature-like amplification (Perelman, 2002).

## Liquidity, Systemic Risk, and Resilience of Financial Transmission (LNSR)

Systemic risk is fundamentally nonlinear: small shocks may trigger disproportionately large system losses through feedback loops, balance-sheet amplification, liquidity spirals, and correlated distress (Brunnermeier & Sannikov, 2014; Gorton, 2010). This insight shifted the literature from micro-prudential soundness to system-wide metrics capturing joint distress and tail risk contributions, including CoVaR and SRISK (Adrian & Brunnermeier, 2016; Brownlees & Engle, 2017).

Network-based analyses of banking ecosystems likewise show that systemic fragility depends not only on individual balance sheets but also on the architecture of interconnections and feedback channels across the financial system (Haldane & May, 2011). Minsky's financial instability hypothesis also remains highly relevant, emphasizing that apparently tranquil periods can generate endogenous fragility through leverage accumulation and deteriorating financing structures (Minsky, 1986). The interaction between market liquidity and funding liquidity is particularly important, as adverse feedback between the two can amplify stress and accelerate systemic deterioration under constrained financing conditions (Brunnermeier & Pedersen, 2009).

Classic contagion theory further demonstrates that localized financial shocks can spread across institutions through interconnected balance sheets and network linkages, making liquidity resilience inseparable from systemic resilience under stress (Allen & Gale, 2000). The financial-cycle literature likewise shows that macroeconomic stability cannot be assessed independently of medium-term credit and asset-price dynamics, which often accumulate beneath standard business-cycle indicators (Borio, 2014). Historical evidence on financial crises further demonstrates that episodes of instability are recurrent, heterogeneous, and often preceded by vulnerabilities that are underestimated in real time (Reinhart & Rogoff, 2009).

Liquidity resilience reflects the capacity to maintain intermediation and payment functionality under stress. Early warning evidence highlights that credit booms, debt burdens, and financial-cycle indicators often predict crises more reliably than contemporaneous macro indicators (Drehmann & Juselius, 2012; Schularick & Taylor, 2012).

Composite stress indices such as the ECB's CISS aggregate market-based signals and emphasize correlation structures under stress, supporting real-time monitoring of systemic conditions (Holló et al., 2012).

Gondauri et al. (2025) advances this literature by modelling liquidity flows and systemic risks through a Navier–Stokes-based economic analogy, where pressure gradients, friction-like terms, and residual forces represent mechanisms of stress propagation and resilience. This methodological direction supports the GI's Liquidity–Navier–Stokes Resilience (LNSR) pillar, which is designed to remain interpretable while compatible with publicly replicable data pipelines.

## Inflation Forecasting Credibility and Forecast Coherence (IFC)

Inflation forecasting is both a technical and an institutional credibility channel. Forecast failures weaken nominal anchors, amplify risk premia, and reduce the effectiveness of stabilization instruments (Faust & Wright, 2013; Stock & Watson, 2007). Consequently, forecasting performance is increasingly treated as a policy-relevant signal rather than an isolated modelling benchmark.

From a forecasting perspective, Bayesian approaches further highlight the importance of model uncertainty, prior information, and probabilistic updating in the evaluation of predictive performance (Geweke & Whiteman, 2006). Regime-dependent dynamics are especially important in macroeconomic forecasting, since nonlinear shifts across latent states can substantially alter the behavior of inflation and other key aggregates (Hamilton, 1989). More generally, modern predictive analytics emphasize that model selection, regularization, and out-of-sample validation are essential for maintaining forecasting discipline when multiple candidate specifications compete for explanatory relevance (Hastie et al., 2009). Comparative forecasting evidence further shows that statistical and machine-learning models each have strengths and limitations, so predictive credibility must be judged empirically rather than assumed from model complexity alone (Makridakis et al., 2018).

Hidden Markov modelling is likewise relevant for regime-sensitive inflation analysis because it provides a formal framework for identifying latent states and transitions in dynamic processes (Rabiner, 1989). Dynamic factor approaches further illustrate how common latent forces can be extracted from large information sets to improve macroeconomic forecasting and strengthen signal extraction under noisy conditions (Stock & Watson, 2011).

Forecasting and Policy Analysis Systems (FPAS) provide a structured platform for combining models, expert judgment, and scenario assessment in monetary-policy decision cycles (Al-Mashat et al., 2018).

However, modern forecasting research shows that no single model dominates across regimes; performance shifts under structural breaks and supply shocks. This motivates coherence-oriented evaluation, where improvements are judged by their stability across windows and regimes rather than one-sample dominance (Clark & McCracken, 2013; Giannone et al., 2008).

Gondauri (2025a) proposes a hybrid FPAS + ζ framework that augments structural FPAS dynamics with a Riemann zeta signal to capture cyclical flexibility and shock responsiveness, supported by Fourier/HMM diagnostics.

### Rolling Diagnostics, Scenario Evidence, and the Binding-Constraint Logic

Operational resilience measurement requires dynamic diagnostics over time, attribution of changes to drivers, and scenario evidence that stress-tests whether stability improves or erodes under alternative macro conditions (IMF, 2019). Rolling-window statistics are particularly valuable in identifying instability before it becomes visible in long-run averages (Drehmann & Juselius, 2012). Contribution decomposition further supports transparent policy narratives by clarifying which indicators drive movements in the composite score (Greco et al., 2018).

A useful decision mechanism is the binding-constraint interpretation: the lowest pillar score identifies the dominant weakness constraining aggregate resilience. This logic is aligned with the non-compensatory nature of resilience and enables actionable sequencing: interventions should target the most fragile channel first to improve system-wide robustness. Work on the financial cycle likewise shows that medium-term credit and asset-price fluctuations contain important early-warning information that may be missed by shorter-horizon macro indicators (Drehmann et al., 2012).

### Research Positioning and the Contribution of the Gondauri Index Framework

This study is explicitly grounded in the methodological framework of the Gondauri Index (GI) (Gondauri, 2025b). Accordingly, the present article should be read as an application built on that GI methodological architecture: the index design, its sub-index logic (IRS, LNSR, IFC), normalization and aggregation principles, and the underlying millennium-problems-driven economic interpretation provide the core analytical scaffold for the empirical results and policy inferences.

The practical feasibility of cross-country resilience benchmarking is strengthened by harmonized international macroeconomic datasets such as the World Development Indicators, which support consistent variable construction across countries and years (World Bank, 2024).

The broader literature highlights persistent gaps that GI addresses through an integrated and reproducible design. First, many monitoring systems remain indicator dashboards without an aggregation rule that reflects the non-substitutability of resilience dimensions, potentially obscuring binding weaknesses (Nardo et al., 2005; OECD, 2023). Second, composite indices sometimes prioritize ranking over interpretability; recent methodological work emphasizes decomposability and robustness as essential properties (Foster et al., 2012; Greco et al., 2018). Third, distributional dynamics are often under-integrated into macro-financial resilience tools despite strong evidence of macro feedbacks (Atkinson, 2015; Coibion et al., 2017). Fourth, systemic risk measures may rely on market-based inputs that reduce feasibility for wide cross-country replication; GI contributes a transparent resilience mapping that can be implemented with publicly available macro series while remaining conceptually consistent with systemic-risk diagnostics (Adrian & Brunnermeier, 2016; Holló et al., 2012). Finally, forecast quality is rarely embedded into resilience indices even though nominal-anchor credibility is central to shock absorption; GI operationalizes this channel through the IFC pillar, informed by FPAS practice and hybrid augmentation evidence (Al-Mashat et al., 2018; Gondauri, 2025a).

## METHODOLOGY

### Research aim

The primary aim of the study is to design and empirically demonstrate a reproducible benchmarking framework that quantifies macro-financial resilience across heterogeneous economies using observable macroeconomic series and transparent diagnostic transformations.

Specifically, the paper introduces the Gondauri Index (GI) and validates it as (i) a comparative resilience scorecard for benchmarking economies and regions and (ii) a diagnostic instrument that attributes resilience dynamics to inequality stability, liquidity/systemic transmission, and nominal-anchor credibility.

### Research objectives and tasks

To operationalize the aim, the study pursues several objectives and implementation tasks. Thus, the methodology proceeds in five structured steps:

Conceptualization of resilience as a structured, measurable construct with economically interpretable dimensions and a consistent "higher is better" 0–100 scoring convention.

Specification of the GI architecture through three pillars (IRS, LNSR, and IFC) and derivation of each pillar from observable macro series using transparent diagnostics. Enforcement of cross-economy comparability through a unified data-vintage policy (latest available ≤ 2024), robust percentile normalization (p5 – p95), and explicit missing-data handling via component-level weight re-normalization. Empirical validation via a 2024 benchmark snapshot, 2005–2024 trajectories with 5-year rolling diagnostics, and Δlog(GI) contribution decomposition for pillar-level attribution.

Extension of the framework to forward-looking intelligence by constructing 2026–2030 scenario pathways and a binding-pillar diagnostic to identify the dominant constraint on resilience across horizons.

### Research questions

This study is guided by four research questions: (i) whether macro-financial resilience can be summarized in a single composite metric without losing interpretability and diagnostic value; (ii) how inequality resilience, liquidity/systemic resilience, and inflation-forecast coherence interact in shaping aggregate resilience; (iii) whether a geometric aggregation rule improves realism by penalizing one-pillar fragility compared to additive composites; and (iv) how stable resilience rankings are over time and what rolling diagnostics reveal about regime shifts.

First, the paper proposes a diagnostics-first composite index—the Gondauri Index (GI)—that summarizes macro-financial resilience on a transparent 0–100 scale while preserving interpretability through pillar-level decomposition and explicit weighting. Second, it quantifies how the three resilience pillars co-move and interact over time by linking inequality resilience (IRS), liquidity/systemic resilience (LNSR), and inflation-forecast coherence (IFC) within a single measurement architecture and by using rolling diagnostics to detect shifts in the relative importance of each pillar. Third, the pillars operationalize resilience using parsimonious and replicable transformations: (i) IRS links inequality dynamics to macro drivers and stability, (ii) LNSR captures stability of liquidity transmission through liquidity-depth proxies and residual-stress diagnostics, and (iii) IFC embeds forecast performance coherence into resilience measurement by comparing benchmark models with a hybrid FPAS + ζ augmentation. Fourth, the paper extends the index into a forward-looking stress-test framework. By mapping scenarios into pillar paths and identifying binding constraints, the GI becomes actionable for policy sequencing rather than a purely descriptive instrument.

### Relevance and importance

Policy relevance is high because central banks and ministries increasingly require resilience metrics that go beyond single-indicator surveillance. The GI provides a structured measure that can support macroprudential monitoring, inflation-targeting credibility assessment, and risk-aware policy sequencing, especially for emerging and frontier economies. Methodologically, the framework prioritizes reproducibility, transparent formulas, and robust scaling, while rolling-window diagnostics and contribution decomposition reduce reliance on single-year snapshots. By embedding inflation-forecast coherence and hybrid signal augmentation (FPAS + ζ), the index also links resilience measurement to predictive credibility and regime sensitivity.

**Scope, dataset, and empirical design**

The empirical implementation is defined over an annual country-year panel. The benchmarking exercise in the Results section uses a selected set of economies over 2005–2024, with a 2024 snapshot and 5-year rolling diagnostics. Macroeconomic indicators are sourced from the World Bank World Development Indicators (WDI), and a uniform vintage policy is applied: where 2024 data are missing, the latest available observation ≤ 2024 is used. Sparse series, particularly inequality (Gini), are handled through explicit weight renormalization at the component level to preserve a well-defined index without excluding country-years.

**Index architecture and identification**

The Gondauri Index (GI) is a composite macro-financial resilience metric on a 0–100 scale. It integrates three pillars that are computed from observable macro series and diagnostic transformations: (i) Inequality Resilience Score (IRS), (ii) Liquidity & Systemic Resilience (LNSR), and (iii) Inflation Forecast Coherence (IFC). The composite is constructed as a weighted geometric aggregation, so that weakness in any single pillar penalizes the overall score.

Pillars and interpretation are the following:
- GI ∈ [0,100], where higher values indicate stronger structural resilience:
- IRS (0–100) – resilience of inequality dynamics (not the welfare level);
- LNSR (0–100) – stability of liquidity transmission and systemic-risk conditions, proxied via liquidity-speed and residual-force diagnostics;
- IFC (0–100) – coherence/credibility of inflation forecasting, quantified by relative forecast error improvements when augmenting FPAS with a zeta-based signal (FPAS + ζ).

*Country-years, scope, and comparability*

The empirical workflow is defined over a country-year panel (annual frequency). As a result, the benchmark set comprises nine selected economies over 2005–2024, with a 2024 cross-sectional snapshot and rolling diagnostics computed using a 5-year moving window. Cross-country comparability is enforced by (i) common variable definitions, (ii) robust percentile scaling, and (iii) consistent rolling-window parameters.

**Data, sources, and preprocessing**

All macro series are sourced from the World Bank World Development Indicators (WDI). For the 2024 snapshot, if an indicator is missing in 2024, the latest available value ≤ 2024 is used (vintage-carry-forward). This policy is applied consistently across all economies and indicators.

**Core indicators and notation**

Let $t$ denote year. The baseline variables are:
- $\pi_t$: CPI inflation rate (%), WDI code FP.CPI.TOTL.ZG;
- $g_t$: real GDP growth rate (%), WDI code NY.GDP.MKTP.KD.ZG;
- $u_t$: unemployment rate (%), WDI code SL.UEM.TOTL.ZS;
- $M3GDP_t$: broad money as % of GDP, WDI code FM.LBL.BMNY.GD.ZS (or the closest broad-money-to-GDP series available);
- $\Delta M3_t$: broad money growth rate (%), derived from broad money levels if available.

**Missing data handling**

Several WDI series are sparse (notably the Gini coefficient). The methodology remains well-defined under missingness by re-normalizing weights within each pillar when a component is unavailable for a given country-year. This avoids dropping country-years and preserves comparability.

*Robust normalization to a 0–100 score*

All raw metrics $m$ are mapped to a common 0–100 scale using robust percentile scaling with 5th–95th percentile bounds computed on the full benchmarking pool (country-year panel). This limits outlier leverage and stabilizes comparisons across heterogeneous economies:

$$\text{Score}_m(x_{m,t}) = \text{clip}_{[0,100]} \times \left(100 \times \frac{x_{m,t} - p5(m)}{p95(m) - p5(m)}\right) \quad (1)$$

where
- $x_{m,t}$ is the raw value of metric m in year $t$;

- $p5(m)$, $p95(m)$ are the 5th and 95th percentiles of metric $m$ computed over the benchmarking pool;
- $\text{clip}_{[0,100]}$ sets negative values to 0 and values above 100 to 100.

Directionality conventions: for 'bad' metrics (e.g., volatility, residual stress, absolute inequality changes), a monotone inversion or sign change is applied before scaling so that higher scores always reflect stronger resilience.

## Pillar 1 — Inequality Resilience Score (IRS)

### Structural inequality model and goodness-of-fit

IRS is anchored in a parsimonious structural regression of inequality (Gini) on macro drivers, estimated separately for each country on the available annual sample. The Results section reports fit diagnostics ($N$, $R^2$, coefficients, and $p$-values) for this baseline specification:

$$\text{Gini}_t = \beta_0 + \beta_\pi \cdot \pi_t + \beta_u \cdot u_t + \beta_y \cdot \log(\text{GDPpc}_t) + \eta_t \quad (2)$$

where

- $\text{GDPpc}_t$ is GDP per capita (current US$) or the most consistent GDP per capita series available in WDI;
- $\eta_t$ is the inequality residual capturing unobserved distributional shocks and measurement noise.

### IRS construction

IRS is operationalized as a weighted aggregation of (i) model explanatory power, (ii) stability of inequality changes, and (iii) a smoothing/structural-stability signal. This matches the GI specification and is consistent with the Results interpretation of 'resilience of dynamics':

$$\text{IRS}_t = 0.40 \cdot \text{Score}(R^2) + 0.40 \times$$
$$\times \text{Score}(\text{Inv} |\Delta\text{Gini}_{\{t,12m\}}|) + 0.20 \cdot \text{Score}(\hat{S}_t), \quad (3)$$

where

- $R^2$ is from Equation (2), estimated on the available country sample;
- $|\Delta\text{Gini}_{\{t,12m\}}|$ denotes an annualized absolute change in Gini (penalized because large swings reduce resilience);
- $\text{Inv}(\cdot)$ is a monotone inversion so that smaller absolute changes map to higher resilience scores;
- $\hat{S}_t$ is a structural smoothing/stability signal (e.g., entropy/Ricci-inspired smoothness proxy);
- if a component is missing (e.g., sparse Gini vintages), the remaining component weights are re-normalized to sum to 1.

### Practical notes

IRS is interpreted as a stability metric, not a welfare rank. Sparse/discontinuous Gini vintages imply that sample size varies by country; this is reported as a data-coverage limitation.

## Pillar 2 — Liquidity & Systemic Resilience (LNSR)

### Liquidity-speed proxy and rolling stability

LNSR quantifies the stability of liquidity transmission using a reduced-form proxy for liquidity speed and a residual-stress diagnostic. Liquidity speed $v_t$ is defined from monetary depth (broad money as a percantage of GDP).

$$v_t = \frac{100}{\text{M3GDP}_t}, \quad (4)$$

where

- $\text{M3GDP}_t$ is broad money as percent of GDP;
- $v_t$ is a liquidity-speed proxy: deeper monetization (higher $\text{M3GDP}_t$) implies lower $v_t$.

Rolling variance (5-year window):

$$\text{Var}_t(v) = \text{Var}(v_{\{t-4\}}, \ldots, v_t), \quad (5)$$

- where $\text{Var}(\ )$ is the sample variance within the rolling window.

### Residual-force proxy for inflation–liquidity balance

The Results implementation defines an inflation residual proxy $\varepsilon_t$ that captures the component of inflation not explained by a reduced-form balance involving a policy/stance term, growth, and liquidity-speed changes:

$$\varepsilon_t = \pi_t - (\mu_t - g_t + \Delta v_t), \quad (6)$$

where

- $\mu_t$ is a policy/monetary stance term (defined consistently within the data environment);

- $g_t$ is real GDP growth;
- $\Delta v_t$ is $v_t - v_{t-1}$.

Rolling RMS (5-year window):

$$\text{RMS}_t(\varepsilon) = \sqrt{\frac{1}{5}\sum_{k=0}^{4}\varepsilon_{t-k}^2}, \quad (7)$$

- where lower RMS implies a more controllable residual component and supports higher resilience.

*Cycle-alignment diagnostic*

$$\text{Align}_t = |\text{Corr}(\Delta v_{t-4:t}, g_{t-4:t})|, \quad (8)$$

- where $\text{Align}_t$ denotes the absolute rolling correlation between liquidity changes and GDP growth.

*LNSR aggregation*

$$\text{LNSR}_t = 0.35 \cdot \text{Score}\left(\frac{1}{1 + \text{Var}_t(v)}\right) + 0.35 \times$$
$$\times \text{Score}\left(\frac{1}{1 + \text{RMS}_t(\varepsilon)}\right) + 0.30 \cdot \text{Score}(\text{Align}_t), \quad (9)$$

where weights are re-normalized if a component is unavailable for a given country-year.

## Pillar 3 — Inflation Forecast Coherence (IFC)

*Forecasting models and rolling evaluation*

IFC evaluates whether inflation forecasting improves when moving from benchmark models to a hybrid structural model augmented with a Riemann zeta-based cyclical signal. The Results section compares AR(1), FPAS-ARX (FPAS baseline), and FPAS + ζ using rolling 5-year RMSE.

$$\text{RMSE}_t(\text{model}) = \sqrt{\frac{1}{H}\sum_{h=0}^{H-1}(\pi_{t-h} - \hat{\pi}_{t-h}^{(\text{model})})^2}, \quad (10)$$

where $H$ is the rolling window length ($H$ = 5 in the Results implementation).

$$\Delta\text{RMSE}_t^{\text{FPAS}\to\text{FPAS}+\zeta} = 100 \cdot \frac{\text{RMSE}_t(\text{FPAS}) - \text{RMSE}_t(\text{FPAS}+\zeta)}{\text{RMSE}_t(\text{FPAS})}, \quad (11)$$

where positive values indicate improvement relative to the FPAS baseline.

$$\Delta\text{RMSE}_t^{\text{AR}\to\text{FPAS}+\zeta} = 100 \times$$
$$\times \frac{\text{RMSE}_t(\text{AR}) - \text{RMSE}_t(\text{FPAS}+\zeta)}{\text{RMSE}_t(\text{AR})}, \quad (12)$$

where positive values indicate improvement relative to AR(1).

Only forecast gains contribute to IFC via truncation:

$$\Delta\text{RMSE}_t^+ = \max(0, \Delta\text{RMSE}_t) \quad (13)$$

where negative improvements are set to zero to prevent 'rewarding' forecast deterioration.

*FPAS + ζ specification and calibration*

$$\hat{\pi}_t^{(\text{FPAS}+\zeta)} = \hat{\pi}_t^{(\text{FPAS})} + \alpha \cdot [\zeta(0.5 + i \cdot t^*) - \bar{\zeta}], \quad (14)$$

where

- $t^*$ is a transformed macro time base aligned with annual frequency;
- ζ is a smoothed reference level (e.g., moving average) used for centering;
- α is calibrated (e.g., by minimizing rolling RMSE on a validation segment).

*Regime diagnostics*

The Results section reports diagnostic filters supporting regime-dependent interpretation:

- regime accuracy: rolling HMM phase classification accuracy on inflation regimes;
- |corr(ζ, Δπ)|: absolute correlation between ζ-signal and inflation changes (alignment proxy).

*IFC aggregation*

$$\text{IFC}_t = 0.35 \cdot \text{Score}(\Delta\text{RMSE}_{\text{FPAS}\to\text{FPAS}+\zeta,t}^+) +$$
$$+0.25 \cdot \text{Score}(\Delta\text{RMSE}_{\text{AR}\to\text{FPAS}+\zeta,t}^+) +$$
$$+0.25 \cdot \text{Score}(\text{Acc}_t^{\text{HMM}}) + 0.15 \cdot \text{Score}(\text{Match}_t^{\zeta-\text{spec}}) \quad (15)$$

where

- $\text{Acc}_t^{\text{HMM}}$ is rolling regime classification accuracy;
- $\text{Match}_t^{\zeta-\text{spec}}$ is a cycle-matching / spectrum-alignment proxy.

## Composite construction of GI: Weighted geometric aggregation

$$\text{GI}_t = (\text{IRS}_t^{w1}\text{LNSR}_t^{w2}\text{IFC}_t^{w3})^{1/(w1+w2+w3)},$$
$$\text{with } (w1, w2, w3) = (0.35, 0.35, 0.30), \quad (16)$$

where geometric aggregation rewards balanced resilience and penalizes single-pillar fragility.

*Contribution decomposition (Δlog GI)*

To attribute year-to-year GI movements to pillar dynamics, we apply the next equation:

$$\Delta \log(\text{GI}_t) = \left(\frac{w1}{W}\right) \cdot \Delta \log(\text{IRS}_t) + \left(\frac{w2}{W}\right) \times \Delta\log(\text{LNSR}_t) + (w3/W) \cdot \Delta\log(\text{IFC}_t), \quad (17)$$

where $W = w1 + w2 + w3$.

*Regional benchmarking protocol*

Regional aggregation follows World Bank region/aggregate codes. When all required series exist for a region aggregate, the identical GI pipeline is applied. If an input is missing for an aggregate, the corresponding pillar is left blank, and the table is used as a benchmarking scaffold (consistent reporting convention).

*Scenario-based projections (2026–2030)*

The Results section includes 2026–2030 scenario-based projections that translate the GI logic into conditional pathways (Baseline/Adverse/Optimistic). This section documents the projection mechanics.

*Baseline interpolation*

$$\text{Pillar}_t^{\text{Base}} = \text{Pillar}_{2026} + \frac{t-2026}{4} (\text{Pillar}_{2030} - \text{Pillar}_{2026}), \quad (18)$$

where $t = 2026, \ldots, 2030$.

Applied separately for IRS, LNSR, and IFC.

*Scenario shock mapping*

$$\text{Pillar}_t^{\text{Scenario}} = \text{clip}_{[0,100]} \left( \text{Pillar}_t^{\text{Base}} + \text{Shock}_t^{\text{Scenario}} \right), \quad (19)$$

where adverse shocks are negative; optimistic shocks are positive; mapping is pillar-symmetric in the Results implementation.

*Binding pillar*

$$\text{BindingPillar}_t = \arg\min\{\text{IRS}_t, \text{LNSR}_t, \text{IFC}_t\}, \quad (20)$$

$$\text{BindingScore}_t = \min\{\text{IRS}_t, \text{LNSR}_t, \text{IFC}_t\} \quad (21)$$

Reported as the annual binding constraint in the scenario pillar table.

## RESULTS

### Terminology and reporting conventions

In this section, a single, consistent notation and a harmonized set of reporting conventions across all tables and figures are applied in order to avoid unit ambiguity and terminology drift.

GI — Gondauri Index (0–100): a weighted geometric aggregation of the three pillars (IRS, LNSR, IFC). Higher values indicate stronger macro-financial resilience;

IRS — Inequality Resilience Score (0–100): a resilience metric for the dynamics of inequality (it does not attempt to measure the welfare level itself);

LNSR — Liquidity & Systemic Resilience (0–100): a stability metric for liquidity transmission and systemic-risk conditions, proxied via velocity/monetary depth and residual-force diagnostics;

IFC — Inflation Forecast Coherence (0–100): a forecast-quality coherence metric capturing relative improvements when moving from baseline FPAS to FPAS + ζ and related benchmark models (rolling evaluation).

Scoring and comparability:

1. robust normalization uses $p_5$–$p_{95}$ bounds (to limit outlier influence);
2. proxy variables are defined as $v_t$ (liquidity-speed proxy) and as $\varepsilon_t$ (inflation residual proxy);
3. rolling windows are 5-year unless stated otherwise;
4. when a component is missing (e.g., sparse Gini vintages), weights are re-normalized so the index remains well-defined.

### 2024 Snapshot

Table 1 provides a cross-sectional comparison for 2024. A higher GI is favorable: it reflects (i) more stable inequality dynamics (IRS), (ii) steadier liquidity-cycle mechanics and smaller residual stress (LNSR), and (iii) higher inflation-forecast credibility (IFC). Cross-country rank should be read jointly with coverage and data vintage.

**Table 1. 2024 Snapshot: GI and pillars (selected economies)**

| Country | GI (0–100) | IRS | LNSR | IFC (0–100) | Inflation (%) | GDP growth (%) | Unemployment (%) | Broad money (% GDP) | Broad money growth (%) | ΔRMSE FPAS→FPAS+Z (%) | ΔRMSE AR→FPAS+ | Regime accuracy |
|---|---|---|---|---|---|---|---|---|---|---|---|---|
| Armenia | 13.90 | 7.90 | 83.24 | 13.93 | 0.27 | 5.90 | 13.33 | 57.79 | 13.66 | 71. | 94. | 0.40 |
| Azerb. | 49.51 |  | 53.80 | 31.73 | 2.21 | 4.07 | 5.59 | 36.80 | 3.15 | 5.87 | 4.62 | 0.80 |
| Azerbaijan | 49.51 |  | 53.80 | 31.73 | 2.21 | 4.07 | 5.59 | 36.80 | 3.15 | 5.87 | 4.62 | 0.80 |
| China | 72.33 | 86.22 | 81.23 | 21.28 | 0.22 | 4.98 | 4.57 | 227.50 | 6.81 | 09. | 82. | 0.80 |
| China | 72.33 | 86.22 | 81.23 | 21.28 | 0.22 | 4.98 | 4.57 | 227.50 | 6.81 | 09. | 82. | 0.80 |
| Georgia | 52.52 | 67.42 | 73.20 | 14.68 | 1.11 | 9.68 | 11.48 | 53.16 | 14.49 | 07. | 13. | 0.60 |
| Georgia | 52.52 | 67.42 | 73.20 | 14.68 | 1.11 | 9.68 | 11.48 | 53.16 | 14.49 | 07. | 13. | 0.60 |
| Romania | 31.78 | 18.80 | 69.62 | 23.55 | 5.72 | 0.92 | 5.38 |  |  | 76. | 79. | 0.60 |
| Russian Fed. | 25.23 | 50.83 |  | 13.94 | 8.43 | 4.34 | 2.53 |  |  | 77. | 16. | 0.60 |
| Türkiye | 28.93 | 21.08 | 30.04 | 45.91 | 58.51 | 3.33 | 8.45 |  |  | 3.16 | 7.40 | 1.00 |
| Ukraine | 43.47 | 63.22 | 55.91 | 14.13 | 6.50 | 2.91 |  | 45.55 | 13.36 | 56. | 91. | 0.60 |
| United States | 41.48 | 29.60 | 77.47 | 21.14 | 2.95 | 2.79 | 4.11 | 100.72 | 5.39 | 91. | 58. | 0.60 |

*Source: author's calculations based on World Bank, World Development Indicators (WDI) (latest available ≤2024), computed in Python (pandas, numpy).*

*Notes: Where 2024 values are missing, the latest available ≤2024 is used. Definitions, units, scoring normalization (p5–p95), and evaluation protocol are provided in Annex A.*

Table 1 indicates a higher GI score for Georgia (52.52) than for the United States (41.48). This ranking is driven primarily by the structure of the index and the use of a multiplicative aggregation across the three pillars (IRS, LNSR, IFC) on a common 0–100 robust scale.

In particular, the United States exhibits a markedly lower IRS (29.60) relative to Georgia (67.42), and this deficit exerts a disproportionate downward effect on the composite GI under a geometric-type aggregation, where a single weak pillar cannot be fully offset by strong performance in other dimensions. While the United States scores higher on LNSR (77.47 vs. 73.20) and IFC (21.14 vs. 14.68), these advantages do not compensate for the comparatively low IRS in the 2024 cross-sectional comparison. Importantly, the IRS should be interpreted as an indicator of inequality resilience (i.e., stability and predictability of inequality dynamics), rather than a statement about the absolute level of inequality.

Therefore, the 2024 ordering reflects pillar-specific relative positioning under robust normalization in that year, not a general claim that the United States underperforms structurally across the full period.

## Regional aggregation (World Bank regions): regional benchmarking

Regional aggregation (Table 2) is introduced as a deliberate benchmarking layer to interpret country GI levels in the context of their macro-regional environment rather than in isolation. Specifically, World Bank regional aggregates are processed through the *same* GI pipeline (using identical data handling rules, normalization, scaling, and aggregation conventions) so that the resulting regional GI is methodologically comparable to each country series. Once all required aggregate inputs are available, this produces a consistent "country vs. region" reference frame that helps distinguish idiosyncratic national movements from broader regional dynamics, strengthens the interpretability of cross-sectional comparisons across the nine selected economies, and provides a transparent contextual anchor for discussing relative performance, divergence, and convergence patterns over time. Robustness/uncertainty note: figures report annual deterministic transformations. Where relevant, practical uncertainty bands can be added (e.g., p5–p95 or p10–p90 bands) to reflect sensitivity to scaling choices and windowing.

Table 2. Regional benchmarking (selected economies as region proxies)

| Aggregate (WB code) | GI (0–100) | IRS | LNSR | IFC | Benchmarking note |
|---|---|---|---|---|---|
| World (WLD) | 39.91 | 43.13 | 65.56 | 22.25 | Proxy: arithmetic mean across selected economies ($N$ = 9). |
| Europe & Central Asia (ECS) | 35.05 | 38.21 | 60.97 | 22.55 | Proxy: mean across selected economies in region ($N$ = 7; IRS $n$ = 6; LNSR $n$ = 6). |
| East Asia & Pacific (EAS) | 72.33 | 86.22 | 81.23 | 21.28 | Proxy: mean across selected economies in region ($N$ = 1; IRS $n$ = 1; LNSR $n$ = 1). |
| Latin America & Caribbean (LCN) | - | - | - | - | Not represented in selected economies; not computed. |
| Middle East & North Africa (MEA) | - | - | - | - | Not represented in selected economies; not computed. |
| North America (NAC) | 41.48 | 29.60 | 77.47 | 21.14 | Proxy: mean across selected economies in region ($N$ = 1; IRS $n$ = 1; LNSR $n$ = 1). |
| South Asia (SAS) | - | - | - | - | Not represented in selected economies; not computed. |
| Sub-Saharan Africa (SSF) | - | - | - | - | Not represented in selected economies; not computed. |
| European Union (EUU) | 31.78 | 18.80 | 69.62 | 23.55 | Proxy: Romania only (sample coverage). |

*Source: author's calculations based on World Bank, World Development Indicators (WDI) regional aggregates (latest available ≤ 2024), computed in Python (pandas, numpy).*

*Notes: Regional aggregates follow World Bank region/aggregate codes (e.g., ECS, EAS, NAC, EUU, WLD). Regional GI requires the same inputs as country GI (including proxies for vt and εt). If an input series is not consistently reported for an aggregate, the corresponding pillar is left blank, and the table is used as a benchmarking scaffold.*

Figure 1 shows that Georgia's GI dynamics are driven primarily by variation in inequality resilience (IRS) and episodes of liquidity-speed stability (LNSR), while the inflation-forecast coherence component (IFC) tends to act as the binding constraint when forecast improvements remain limited. Missing observations are handled through component-level weight re-normalization. Robustness/uncertainty note: All series are annual and use deterministic transformations with bounded scaling. GI and pillar components are scaled using p5–p95 bounds (winsorized at the 5th/95th percentiles) to reduce sensitivity to extreme observations; results are stable to alternative reasonable bounds and window lengths.

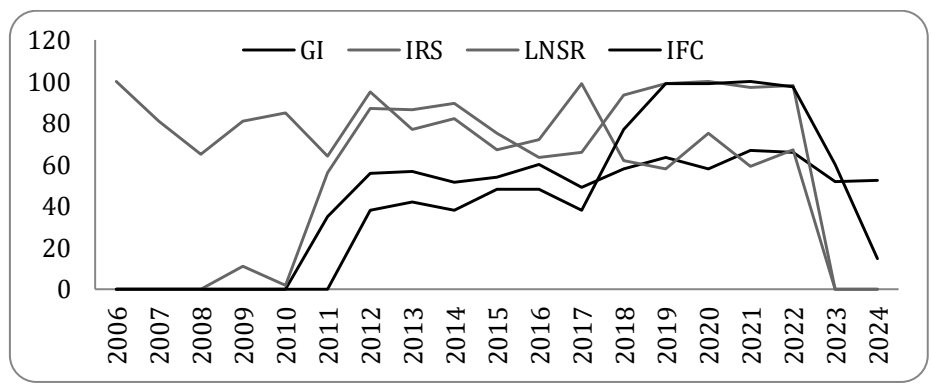

**Figure 1. Georgia: Gondauri Index (GI) and pillar dynamics (2005–2024)**

*Source: Author's calculations based on World Bank, World Development Indicators (WDI) annual series (data vintage up to year 2024), computed in Python (pandas, numpy).*

Figure 1 summarizes the co-movement of the composite GI and its pillars. In Georgia, the GI profile is broadly consistent with shifts in inequality resilience (IRS) and liquidity/systemic resilience (LNSR), while inflation-forecast coherence (IFC) can act as a limiting component in periods when forecast performance gains are modest. Missing pillar observations are handled by renormalizing weights over available pillars (i.e., weights are rescaled to sum to 1 across non-missing pillars in a given year), preserving comparability of the composite index.

Figure 2 shows GI trajectories across the selected economies for the period 2005–2024. Figure 2 should be interpreted as a structural macro-financial resilience comparison across countries, rather than as a welfare ranking. The cross-country GI trajectories reflect differences in the index's diagnostic pillars (IRS, LNSR, IFC) under a common 0–100 scaling. As a robustness check, the main cross-country ordering is stable to alternative bounded-scaling choices (e.g., p5–p95 versus p10–p90) and to the inclusion of supplementary welfare controls (used only for sensitivity, not for index construction).

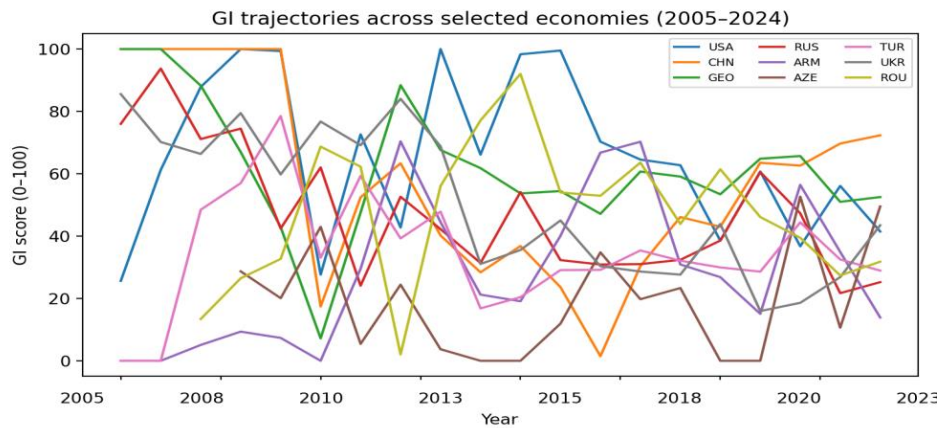

**Figure 2. GI trajectories across the selected economies (2005–2024)**

*Source: author's calculations based on World Bank, World Development Indicators (WDI) (latest available ≤2024), computed in Python (pandas, numpy).*

Figure 3 illustrates the decomposition of the Gondauri Index (GI) into its three diagnostic pillars (IRS, LNSR, and IFC) highlighting how pillar-level movements map into changes in the composite 0–100 resilience score over the sample period. Higher values indicate stronger resilience; pillar components are scaled using bounded percentiles to limit sensitivity to extremes.

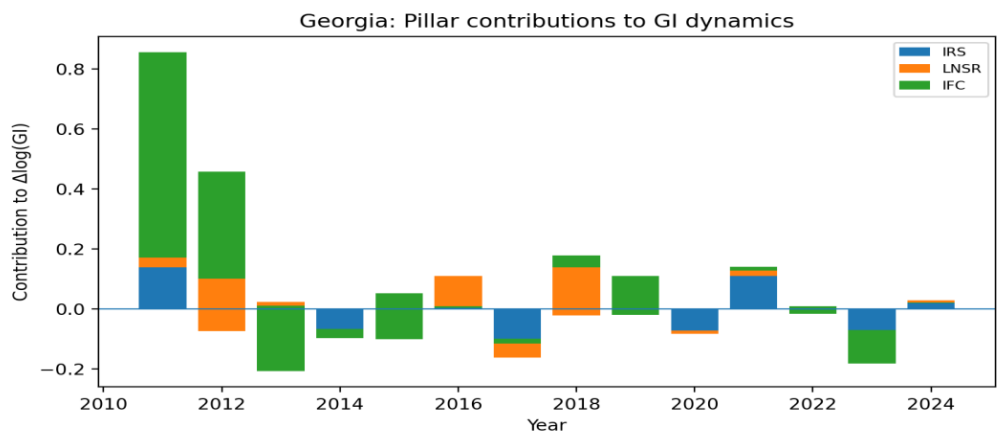

**Figure 3. Georgia: Pillar contributions to Δlog(GI)**

*Source: author's calculations based on World Bank, World Development Indicators (WDI) (latest available ≤ 2024), computed in Python (pandas, numpy).*
*Note: The Δlog(GI) decomposition is derived by log-differentiating the weighted geometric aggregation formula for GI.*

The decomposition in Figure 3 identifies which pillar drives year-to-year changes in the GI, helping diagnose whether movements are primarily associated with inequality resilience (IRS), liquidity/systemic stability (LNSR), or inflation-forecast coherence (IFC). Robustness checks using alternative rolling-window lengths and bounded-scaling choices yield qualitatively consistent pillar-contribution patterns.

Figure 4 shows how robust bounded scaling (p5/p95) limits the leverage of extreme observations in the full country-year panel, improving cross-country comparability of the illustrative metrics used to construct pillar scores and the GI.

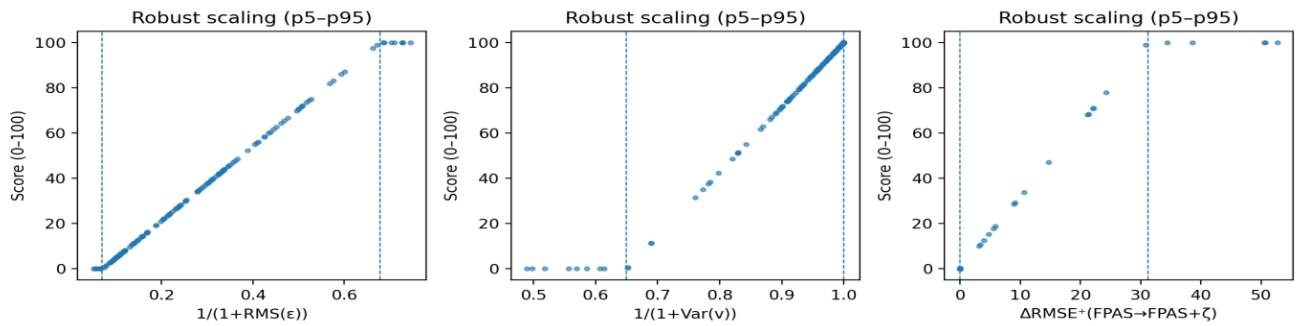

**Figure 4. Robust scaling maps (illustrative metrics)**

*Source: author's calculations based on World Bank, World Development Indicators (WDI) (latest available ≤ 2024), computed in Python (pandas, numpy).*

*Note: The p5/p95 bounds are computed on the full country-year panel (9 countries × 20 years).*

Robust bounded scaling improves cross-country comparability by limiting the influence of extreme observations while preserving relative variation in the central range. The cross-country patterns remain qualitatively stable under alternative percentile bounds (e.g., p10–p90 instead of p5–p95) and under reasonable changes in rolling-window length, indicating that the results are not driven by a single scaling or windowing choice.

**Pillar diagnostics: IRS, LNSR, and IFC**

Figure 5 proves that IRS is higher when inequality dynamics are more stable and less shock-sensitive (e.g., smaller adverse swings in the inequality proxy and stronger alignment with the macro drivers used in the IRS fit). The IRS does not 'explain' the inequality level; it measures the resilience of the dynamics.

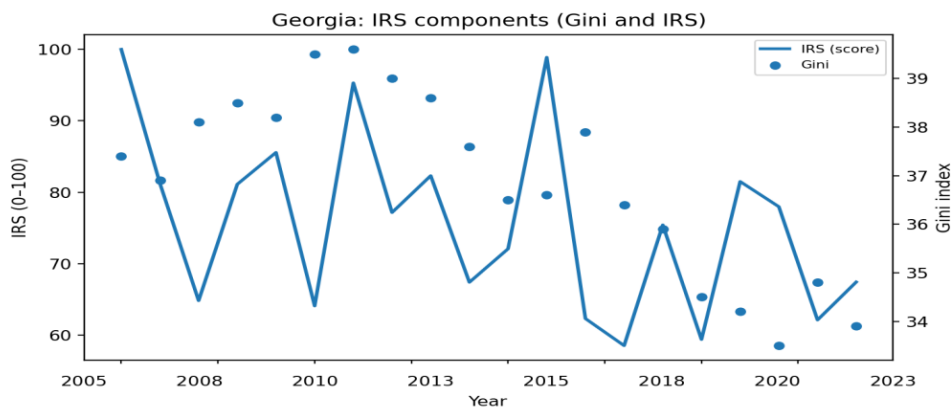

**Figure 5. Georgia: IRS block diagnostics (Gini and IRS)**

*Source: Author's calculations based on World Bank, World Development Indicators (WDI) (latest available ≤ 2024), computed in Python (pandas, numpy).*

*Note: The WDI Gini series is sparse and discontinuous. In IRS construction, component weights are renormalized whenever a sub-component is not observable for a given country-year.*

Table 3 summarizes the 2005–2024 distribution of the Gondauri Index (GI) for each economy – reporting the mean, dispersion (standard deviation), and extrema (minimum/maximum). The table supports both long-horizon comparison and contextual interpretation of the 2024 snapshot. Within this sample and under the adopted bounded scaling, Romania and the United States record the highest average GI, while China remains above the sample mean. Larger standard deviations (e.g., China ≈30.8, USA ≈25.3, Ukraine ≈23.2) indicate substantial time variation consistent with regime changes; therefore, single-year readings are best interpreted alongside 3–5-year rolling averages. Very low minima (including values near zero for some economies) are a mechanical implication of bounded scaling and the 0–100 mapping in periods where one or more pillars approach their lower bounds, while maxima near 100 reflect proximity to the upper bounds under percentile-based scaling. For interpretive clarity, the narrative emphasizes score bands (0–25; 25–50; 50–75; 75–100) rather than fine-grained rank differences.

**Table 3. Descriptive statistics of GI (0–100), 2005–2024**

| Country | Mean GI (2005–2024) | Std | Min | Max | GI 2024 |
|---|---|---|---|---|---|
| China | 57.57 | 30.83 | 1.47 | 100.00 | 72.33 |
| Georgia | 61.66 | 21.24 | 7.19 | 100.00 | 52.52 |
| Azerbaijan | 19.29 | 17.70 | 0.00 | 52.65 | 49.51 |
| Ukraine | 50.35 | 23.16 | 15.96 | 85.54 | 43.47 |
| United States | 65.61 | 25.28 | 25.64 | 100.00 | 41.48 |
| Romania | 47.30 | 22.69 | 2.04 | 92.04 | 31.78 |
| Türkiye | 34.54 | 18.64 | 0.00 | 78.50 | 28.93 |
| Russian Federation | 47.21 | 20.26 | 21.70 | 93.73 | 25.23 |
| Armenia | 28.08 | 23.52 | 0.00 | 70.38 | 13.90 |

*Source: Author's calculations based on World Bank, World Development Indicators (WDI) annual series (data vintage up to year 2024), computed in Python (pandas, numpy).*

Table 4 reports the regression-fit diagnostics for the 'structural' component of IRS, estimated on the available sample window (N years) with a parsimonious specification such as Ginit ~ πt + ut + log(GDPpct). High $R^2$ values (e.g., Georgia ~0.79) support internal validity: the pillar is not ad hoc but anchored in observable macro drivers; however, coefficients should be interpreted with the usual caveats given sample length and data gaps. Coefficient signs vary across countries, reflecting structural differences. In some cases, the inflation coefficient is not statistically significant, which can indicate compensating policy channels and/or measurement noise. p-values are critical: if p(π) or p(u) is high, the fitted inequality dynamics may be driven primarily by labor-market conditions and income levels rather than by inflation.

**Table 4. Inequality model fit used in IRS (Gini ~ inflation + unemployment + log GDPpc).**

| Country | N | $R^2$ | β(π) | β(u) | β(log GDPpc) | p(π) | p(u) | p(log GDPpc) |
|---|---|---|---|---|---|---|---|---|
| Georgia | 20 | 0.791 | 003. | 0.408 | 467. | 0.970 | 0.000 | 0.031 |
| China | 15 | 0.781 | 0.823 | 135. | 335. | 0.011 | 0.160 | 0.010 |
| Ukraine | 16 | 0.598 | 023. | 919. | 344. | 0.460 | 0.005 | 0.008 |
| Russian Federation | 19 | 0.569 | 0.128 | 1.919 | 2.088 | 0.390 | 0.002 | 0.360 |
| United States | 19 | 0.415 | 042. | 236. | 781. | 0.599 | 0.006 | 0.370 |
| Armenia | 19 | 0.325 | 249. | 0.022 | 348. | 0.210 | 0.907 | 0.057 |
| Türkiye | 18 | 0.297 | 0.070 | 0.258 | 868. | 0.039 | 0.499 | 0.812 |
| Romania | 19 | 0.267 | 280. | 0.267 | 143. | 0.074 | 0.709 | 0.618 |

*Source: author's calculations based on World Bank, World Development Indicators (WDI) (latest available ≤ 2024), computed in Python (pandas, numpy).*
*Note: Only countries with sufficient SI.POV.GINI observations are reported (N ≥ 8). For Azerbaijan, IRS is sensitive to sparse Gini vintages and any interpolation/vintage choices.*

Figure 6 shows the evolution of the liquidity-speed proxy v_t = 100 / (Broad money as % of GDP) and its rolling variance Var(v) for Georgia, illustrating periods of stable vs. volatile liquidity transmission.

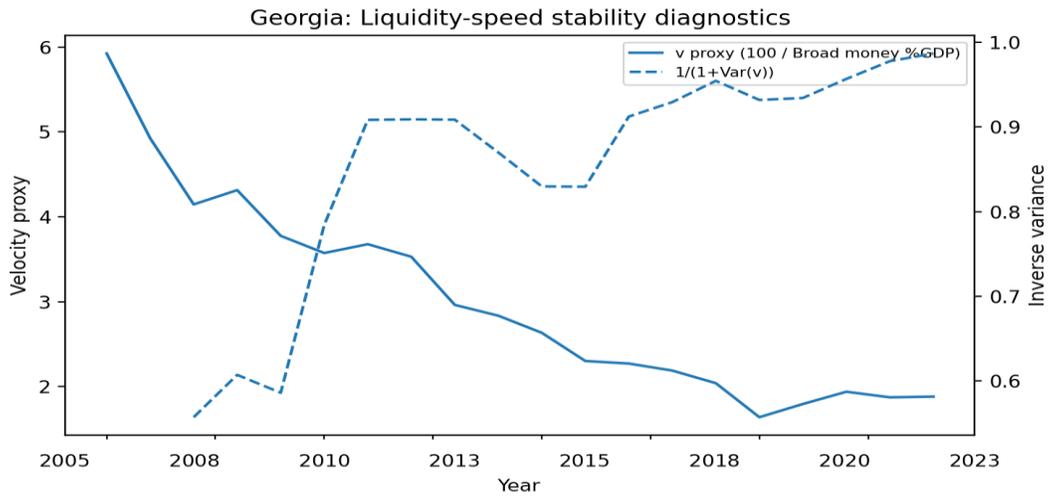

**Figure 6. Georgia: v-proxy and rolling stability (Var(v)).**

*Source: author's calculations based on World Bank, World Development Indicators (WDI) (latest available ≤ 2024), computed in Python (pandas, numpy).*

*Note: vt = 100 / (Broad money as % of GDP). Var(v) is computed in a 5-year rolling window.*

In Figure 6 framework, stability of vt (and of residual diagnostics in the liquidity transmission channel) signals lower systemic-risk vulnerability. When monetary depth and transmission become unstable, the risk of stress increases and LNSR declines.

Figure 7 illustrates the evolution of the residual proxy $\varepsilon_t = \pi_t - (\mu_t - g_t + \Delta v_t)$ and its 5-year rolling stability measure RMS(ε) for Georgia, highlighting periods when inflation is more (or less) aligned with the reduced-form structural relation. As seen from Figure 7, ε t captures the component of inflation not explained by the reduced-form structural relation. Lower RMS(ε) indicates a more controllable 'residual force' and supports higher LNSR / better coherence conditions.

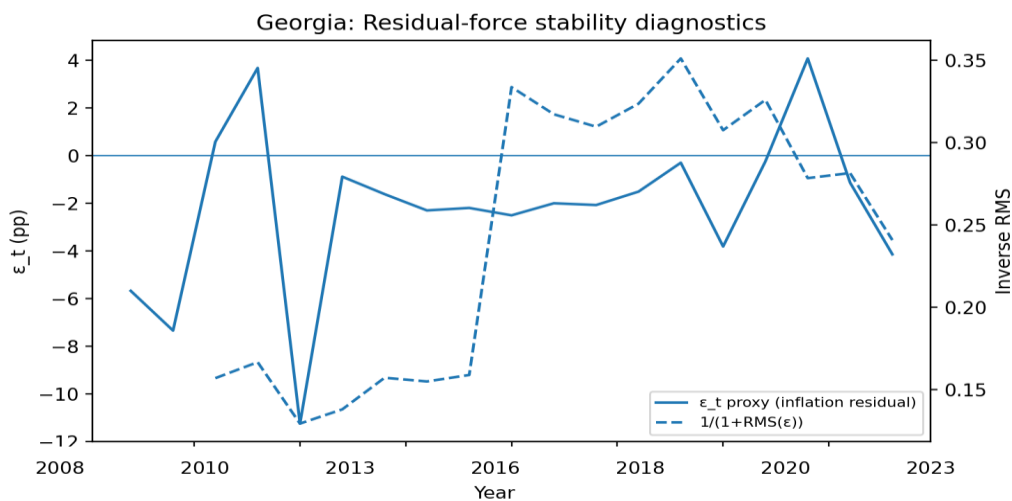

**Figure 7. Georgia: εt proxy and rolling stability (RMS(ε)).**

*Source: author's calculations based on World Bank, World Development Indicators (WDI) (latest available ≤ 2024), computed in Python (pandas, numpy).*

*Note: εt = π t − (μ t − g t + Δv t). RMS(ε) is computed in a 5-year rolling window.*

Figure 8 shows the evolution of the cycle-alignment proxy |corr(Δv, GDP growth)| for Georgia, summarizing how closely liquidity dynamics co-move with real activity over time. As seen from Figure 8, the cycle-alignment component captures the degree to which liquidity dynamics co-move with real activity. Higher alignment is consistent with more coherent transmission, while lower alignment indicates decoupling that is more likely to be driven by shocks or structural breaks.

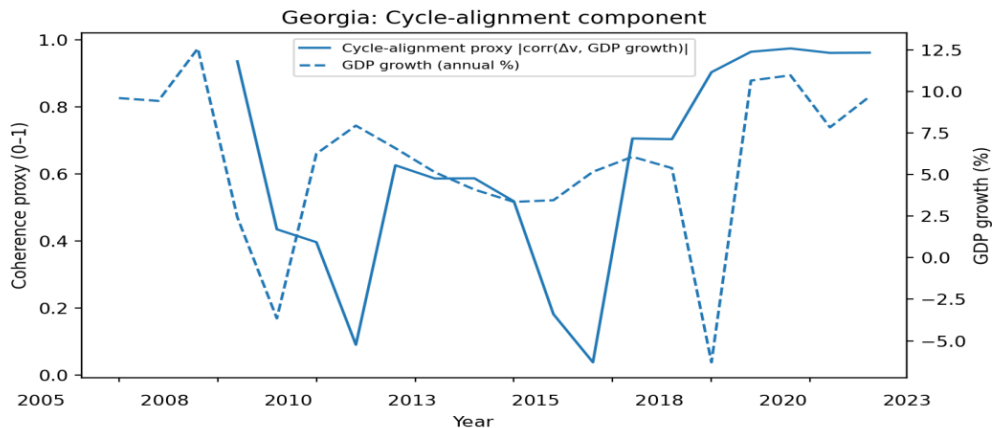

**Figure 8. Georgia: Cycle-alignment proxy (|corr(Δv, GDP growth)|).**

*Source: author's calculations based on World Bank, World Development Indicators (WDI) annual series (data vintage up to year 2024), computed in Python (pandas, numpy).*

*Note: The proxy is computed as the absolute value of the Pearson correlation between Δv_t and GDP growth, estimated in a 5-year rolling window (annual data).*

## Inflation forecasting and the IFC metric

Figure 9 depicts actual inflation versus model-based forecasts for Georgia (AR(1), FPAS-ARX, and FPAS+ζ), enabling a visual assessment of forecast tracking and error dynamics over time. It proves that FPAS + ζ is implemented as an FPAS-ARX baseline augmented with a ζ-signal term. Improvements are recorded using $\Delta RMSE^+ = \max(0, \Delta RMSE)$ so that only forecast gains contribute to IFC.

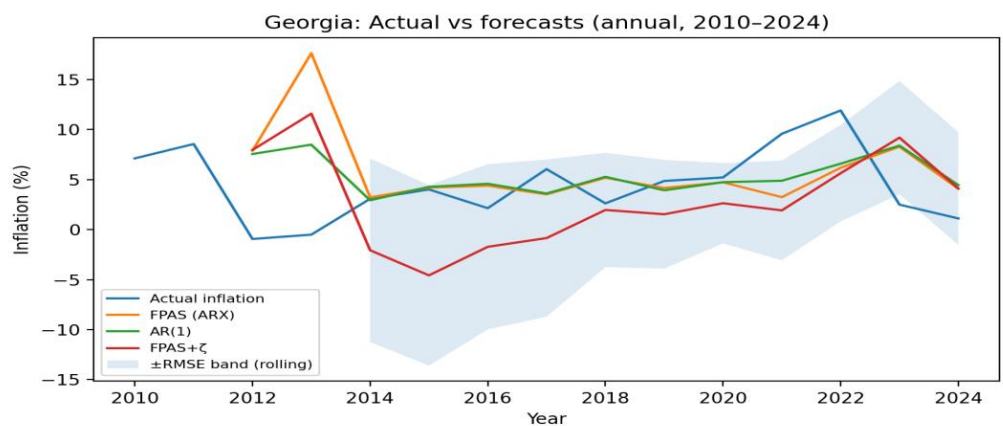

**Figure 9. Georgia: Actual vs forecasts (AR(1), FPAS-ARX, FPAS+ζ).**

*Source: author's calculations based on World Bank, World Development Indicators (WDI) annual series (data vintage up to year 2024), computed in Python (pandas, numpy).*

*Note: Rolling RMSE is computed using a 5-year moving window on one-step-ahead forecast errors; for each year ttt, RMSE is calculated over observations (t−4,…,t)(t-4,\dots,t)(t−4,…,t) for each model on the same sample.*

Table 5 provides the core evidence for IFC: it compares forecast error (RMSE) across models (AR(1), FPAS, FPAS + ζ). Lower RMSE indicates better forecast performance on the same scale. ΔRMSE indicators are interpreted as an incremental gain from adding ζ. Negative or near-zero gains imply that ζ does not improve performance under the current calibration and may require regime-specific tuning.

**Table 5. Forecast performance matrix (2024, rolling metrics).**

| Country | RMSE AR(1) | RMSE FPAS-ARX | RMSE FPAS+ζ | ΔRMSE AR→FPAS+ζ (%) | ΔRMSE FPAS→FPAS+ζ (%) | Regime accuracy (rolling) | \|corr(ζ, Δπ)\| |
|---|---|---|---|---|---|---|---|
| Azerbaijan | 4.288 | 4.345 | 4.090 | 4.625 | 5.868 | 0.800 | 0.050 |
| Türkiye | 73.500 | 70.279 | 68.058 | 7.404 | 3.160 | 1.000 | 0.278 |
| Ukraine | 6.496 | 8.989 | 9.219 | 914. | 558. | 0.600 | 0.084 |
| United States | 3.037 | 3.268 | 3.722 | 578. | 914. | 0.600 | 0.300 |
| Romania | 4.536 | 4.577 | 5.252 | 787. | 759. | 0.600 | 0.374 |
| Georgia | 4.390 | 4.805 | 5.625 | 125. | 072. | 0.600 | 0.101 |
| Russian Federation | 3.884 | 3.426 | 4.240 | 155. | 765. | 0.600 | 0.078 |
| Armenia | 3.551 | 3.204 | 4.508 | 935. | 712. | 0.400 | 0.270 |
| China | 1.649 | 1.157 | 1.794 | 815. | 088. | 0.800 | 0.112 |

*Source: calculated using Stata.*

In this sample, a clear positive signal appears for Azerbaijan, while some economies exhibit mixed regimes in which the ζ-signal can become out-of-phase. For Turkey, RMSE levels are comparatively high; robustness checks should consider alternative error metrics (e.g., MAPE/MASE) or modeling inflation in differences (Δπ) depending on stationarity conditions. Regime accuracy and |corr(ζ, Δπ)| can be used as diagnostic filters to identify windows in which the ζ-signal is empirically aligned and adds predictive power.

*Key findings*

1) GI provides a structural resilience metric that integrates inequality resilience (IRS), liquidity & systemic stability (LNSR), and inflation-forecast coherence (IFC) into a single 0–100 score.

2) Data frequency (annual) and discontinuous Gini vintages introduce unavoidable measurement uncertainty; weight re-normalization mitigates missingness, but results should be interpreted with the stated robustness bounds.

3) The FPAS + ζ effect is heterogeneous across countries; in practice, the ζ module can improve forecasts in some regimes and degrade them in others unless regime-switching calibration is applied.

4) Variant A is designed as a minimal, transparent, and reproducible package. Robustness has been assessed using alternative bounded-scaling choices and reasonable variations in rolling-window length; where higher-frequency data become available, the same diagnostics can be replicated at quarterly (or monthly) frequency as an extension.

Forward-looking outlook: 2026–2030 (scenario-based interpretation). This section provides a forward-looking assessment. It is not a deterministic 'forecast'; rather, it translates the GI logic into scenario-conditioned narratives for 2026–2030 to stress-test how pillars might evolve under alternative macro assumptions.

Outlook matrix: qualitative regional benchmarking and pillar drivers. Table 6 presents a 2026–2030 scenario-based outlook matrix that summarizes the expected qualitative direction of the Gondauri Index (GI) and the dominant pillar-level drivers for each economy. The matrix links a stylized macro backdrop to anticipated movements in GI (rising / stable / declining / volatile), identifies the most likely binding pillars (IRS, LNSR, IFC), and highlights key risks or shocks that could shift trajectories. The entries are intended as structured scenario narratives rather than quantitative point forecasts.

Table 6. 2026–2030 Outlook matrix: qualitative GI direction and pillar drivers (scenario-based)

| Country | Expected macro backdrop (2026–2030) | GI direction | Main drivers (pillars) | Key risks/shocks |
|---|---|---|---|---|
| Georgia | Inflation remains in a low/stable range with moderate growth; gradual deepening of financial intermediation. | Neutral → moderately rising | LNSR: liquidity-speed stability; IFC: forecast discipline | FX shocks; energy/commodity prices; regional security |
| United States | Inflation converges to target; growth normalizes; deep and liquid markets remain supportive. | Stable → slightly rising | IFC: forecast coherence; LNSR: market depth | Fiscal imbalance; financial-cycle risk |
| China | Structural growth slowdown; low inflation; credit rebalancing and policy-driven stabilization. | Slightly declining / volatile | LNSR: credit/financial stability; IRS: distributional pressures | Banking/real-estate stress; geo-economic tensions |
| Russia | High inflation volatility under sanctions and constrained external financing/trade. | Declining / volatile | IFC: forecast coherence under volatility; LNSR: financial isolation effects | Sanctions/financial isolation; energy-price swings; war-related shocks |
| Armenia | Cyclical growth with strong reliance on transfers and trade/logistics corridors. | Neutral / volatile | LNSR: inflow cycles; IFC: forecast discipline | Regional logistics disruptions; financial shocks |
| Azerbaijan | Oil-revenue dependence; inflation and liquidity sensitive to FX and commodity cycles. | Neutral → improving if disinflation holds | IFC: forecast gains; LNSR: revenue/liquidity cycle; IRS | Oil-price shocks; fiscal-cycle volatility |
| Turkey | Disinflation path depends on policy credibility and regime consistency; high FX sensitivity. | Improving under credible disinflation; otherwise volatile | IFC: forecast gains with disinflation; LNSR: FX/liquidity stability | Policy inconsistency; FX shocks; external financing stress |
| Ukraine | Reconstruction phase with high uncertainty; reliance on external financing and support. | Gradual recovery but fragile | LNSR: systemic stability; IRS: social dynamics | Security risk; continuity of fiscal/financial support |
| Romania | EU convergence trajectory; inflation normalizes with steady growth and institutional anchoring. | Moderately high and stable | IFC / LNSR | Energy-price shocks; regional spillovers |

*Source: author's calculations based on World Bank, World Development Indicators (WDI) (latest available ≤ 2024), computed in Python (pandas, numpy).*

*Quantitative projections (three-scenario framework): GI, bands, and pillar paths*

The 2026–2030 values reported below are scenario-based projections anchored in the index structure. They should be read as conditional pathways (Baseline/Adverse/Optimistic), not as point forecasts. Table 7 provides scenario-based GI projections for 2026–2030 on the 0–100 scale, reported alongside an interpretive band classification for each year. For each economy, the table presents conditional paths under Baseline, Adverse, and Optimistic assumptions, allowing readers to compare both the level of projected resilience and the implied transitions across bands over time. These projections are intended as structured stress-test pathways anchored in the GI architecture rather than as unconditional point forecasts.

**Table 7. Scenario projections of GI (0–100) with band classification, 2026–2030 (selected economies)**

| Country | Scenario | 2026 | 2027 | 2028 | 2029 | 2030 |
|---|---|---|---|---|---|---|
| Georgia | Baseline | 54.73 (50–75) | 55.82 (50–75) | 56.89 (50–75) | 57.95 (50–75) | 59.00 (50–75) |
| Georgia | Adverse | 50.73 (50–75) | 50.82 (50–75) | 50.89 (50–75) | 51.45 (50–75) | 52.00 (50–75) |
| Georgia | Optimistic | 58.73 (50–75) | 60.32 (50–75) | 61.89 (50–75) | 63.45 (50–75) | 65.00 (50–75) |
| USA | Baseline | 41.29 (25–50) | 41.19 (25–50) | 41.09 (25–50) | 41.00 (25–50) | 40.90 (25–50) |
| USA | Adverse | 37.29 (25–50) | 36.19 (25–50) | 35.09 (25–50) | 34.50 (25–50) | 33.90 (25–50) |
| USA | Optimistic | 45.29 (25–50) | 45.69 (25–50) | 46.09 (25–50) | 46.50 (25–50) | 46.90 (25–50) |
| China | Baseline | 71.74 (50–75) | 71.44 (50–75) | 71.15 (50–75) | 70.85 (50–75) | 70.55 (50–75) |
| China | Adverse | 67.74 (50–75) | 66.44 (50–75) | 65.15 (50–75) | 64.35 (50–75) | 63.55 (50–75) |
| China | Optimistic | 75.74 (75–100) | 75.94 (75–100) | 76.15 (75–100) | 76.35 (75–100) | 76.55 (75–100) |
| Russia | Baseline | 24.06 (0–25) | 23.46 (0–25) | 22.86 (0–25) | 22.25 (0–25) | 21.63 (0–25) |
| Russia | Adverse | 20.06 (0–25) | 18.46 (0–25) | 16.86 (0–25) | 15.75 (0–25) | 14.63 (0–25) |
| Russia | Optimistic | 28.06 (25–50) | 27.96 (25–50) | 27.86 (25–50) | 27.75 (25–50) | 27.63 (25–50) |
| Armenia | Baseline | 14.41 (0–25) | 14.66 (0–25) | 14.90 (0–25) | 15.15 (0–25) | 15.39 (0–25) |
| Armenia | Adverse | 10.41 (0–25) | 9.66 (0–25) | 8.90 (0–25) | 8.65 (0–25) | 8.39 (0–25) |
| Armenia | Optimistic | 18.41 (0–25) | 19.16 (0–25) | 19.90 (0–25) | 20.65 (0–25) | 21.39 (0–25) |
| Azerb. | Baseline | 49.72 (25–50) | 49.82 (25–50) | 49.93 (25–50) | 50.03 (50–75) | 50.14 (50–75) |
| Azerb. | Adverse | 45.72 (25–50) | 44.82 (25–50) | 43.93 (25–50) | 43.53 (25–50) | 43.14 (25–50) |
| Azerb. | Optimistic | 53.72 (50–75) | 54.32 (50–75) | 54.93 (50–75) | 55.53 (50–75) | 56.14 (50–75) |
| Turkey | Baseline | 29.98 (25–50) | 30.51 (25–50) | 31.03 (25–50) | 31.55 (25–50) | 32.08 (25–50) |
| Turkey | Adverse | 25.98 (25–50) | 25.51 (25–50) | 25.03 (25–50) | 25.05 (25–50) | 25.08 (25–50) |
| Turkey | Optimistic | 33.98 (25–50) | 35.01 (25–50) | 36.03 (25–50) | 37.05 (25–50) | 38.08 (25–50) |
| Ukraine | Baseline | 46.09 (25–50) | 47.36 (25–50) | 48.61 (25–50) | 49.84 (25–50) | 51.06 (50–75) |
| Ukraine | Adverse | 42.09 (25–50) | 42.36 (25–50) | 42.61 (25–50) | 43.34 (25–50) | 44.06 (25–50) |
| Ukraine | Optimistic | 50.09 (50–75) | 51.86 (50–75) | 53.61 (50–75) | 55.34 (50–75) | 57.06 (50–75) |
| Romania | Baseline | 32.16 (25–50) | 32.35 (25–50) | 32.54 (25–50) | 32.73 (25–50) | 32.92 (25–50) |
| Romania | Adverse | 28.16 (25–50) | 27.35 (25–50) | 26.54 (25–50) | 26.23 (25–50) | 25.92 (25–50) |
| Romania | Optimistic | 36.16 (25–50) | 36.85 (25–50) | 37.54 (25–50) | 38.23 (25–50) | 38.92 (25–50) |

*Source: author's calculations based on World Bank, World Development Indicators (WDI) (latest available ≤ 2024), computed in Python (pandas, numpy).*

*Notes: Values are scenario-based projections for 2026–2030 built on the baseline path and stylized shocks. Band classification uses fixed thresholds: 0–25 (very low), 25–50 (low), 50–75 (moderate), 75–100 (high). Inter-country comparability is maintained by the same scaling convention as in Table 1.*

Table 7 consolidates three scenario pathways (Baseline, Adverse, and Optimistic) and the corresponding annual GI band classifications for 2026–2030, reported alongside the underlying pillar scores to preserve interpretability at the component level. The values are produced from the author's computations based on World Bank World Development Indicators (WDI) (latest available series up to ≤ 2024) and implemented in Python (pandas, numpy). Importantly, the 2026–2030 outputs are conditional and index-consistent within the GI framework; they are intended to support structured comparison and policy discussion rather than to serve as definitive point forecasts.

Table 1A (Appendix 1) shows the scenario-based decomposition of the GI into its three pillars (IRS, LNSR, IFC) for 2026–2030 and identifies, for each year, the binding constraint (the lowest pillar score) that limits the composite resilience outcome. By reporting the pillar scores alongside the GI level and band

classification under Baseline, Adverse, and Optimistic pathways, the table clarifies whether projected movements are primarily constrained by inequality resilience, liquidity/systemic stability, or inflation-forecast coherence, and it provides the corresponding binding score for transparent diagnostics.

Table 1A (Appendix 1) links the scenario pathways at the pillar level (IRS, LNSR, IFC) to the binding constraint (the "binding pillar") that caps GI under each scenario. At the country level, the binding mechanism remains interpretable: for example, in Georgia and Turkey, IFC may continue to bind if forecast gains are limited; in Ukraine, LNSR can become binding under adverse liquidity-stress conditions; and in Romania, IRS may bind if inequality dynamics deteriorate.

*How to read GI, and why values may differ across sources*

**Directionality:** GI is reported on a 0–100 scale; higher values indicate higher macro-financial resilience/stability and stronger forecast coherence.

**Bands (practical interpretation):** 0–40 = low resilience; 40–60 = moderate; 60–80 = high; 80–100 = very high. These bands are interpretive aids rather than regulatory classifications.

**Primary sources of differences across releases:** (i) data vintage and subsequent revisions; (ii) indicator definitions and proxy choices; (iii) bounded-scaling conventions (e.g., p5–p95 versus p10–p90); and (iv) horizon/window choices (annual versus quarterly frequency; rolling-window length).

**Reproducibility note:** To ensure exact replication, the Results package documents dataset versions and download timestamps, the code identifier (hash/version tag), rolling-window lengths, and the pillar weights used in aggregation.

## DISCUSSION

The empirical results presented in this manuscript indicate that macro-financial resilience should be interpreted as an internally structured capacity rather than as a single observed outcome. This distinction is central to the contribution of the Gondauri Index (GI). The paper does not merely assemble a list of macroeconomic variables into a synthetic score; instead, it operationalizes resilience as a configuration of three interacting dimensions: the stability of inequality dynamics, the coherence of liquidity and systemic transmission, and the credibility of inflation forecasting. Read in this way, the index contributes to the resilience literature by shifting the analytical focus from descriptive monitoring to disciplined diagnosis. Such a shift is important because economies do not usually become fragile through the simultaneous collapse of all dimensions. More commonly, pressure accumulates through one weak channel that remains partially hidden by apparently favorable performance elsewhere. A meaningful composite framework therefore has to identify weak links rather than average away underlying asymmetries.

Within this logic, the present findings strongly support the proposition that resilience is multi-dimensional and only partially substitutable. The 2024 snapshot, the rolling pillar diagnostics, and the scenario-based projections all point in the same direction: overall resilience depends not simply on the existence of strengths, but on whether those strengths are sufficiently balanced across the essential domains represented by IRS, LNSR, and IFC. This point matters both conceptually and empirically. Conceptually, it challenges highly compensatory measurement approaches in which strong performance in one area can fully offset structural fragility in another. Empirically, it helps explain why some economies with deep financial systems or relatively stable headline inflation do not automatically occupy the highest positions in the composite ranking. The GI framework therefore delivers a more disciplined interpretation of resilience, one in which imbalance is not treated as an innocent statistical feature but as a genuine source of vulnerability.

A key implication of the results is that the geometric aggregation rule serves as an economic mechanism, not merely as a mathematical convenience. In additive indices, pillar strengths can offset pillar weaknesses in a way that risks underestimating fragility. By contrast, the GI construction deliberately penalizes one-pillar weakness. This is consistent with the core claim of the manuscript that resilience failures usually emerge from bottlenecks, not from simple averages. The practical effect of this choice is visible in the cross-country comparison, where the relative position of an economy depends not only on whether one dimension is strong, but also on whether the remaining dimensions are sufficiently robust to sustain an integrated resilience profile. In this respect, the index is closer to the logic of system reliability than to conventional ranking frameworks: a weak link constrains the performance of the whole architecture. For policy interpretation, this is an advantage, because it generates a more realistic picture of where intervention is actually needed.

The behavior of the IRS pillar is especially important in this framework because it clarifies a distinction that is often blurred in comparative macroeconomic analysis. The paper does not equate inequality resilience with the level of inequality itself. Rather, IRS captures the stability, smoothness, and macroeconomic explainability of distributional change. This is a much more specific and analytically defensible object than a generic inequality score. The results imply that distributional dynamics matter for resilience not only because inequality is socially salient, but because instability in inequality can weaken macro-financial adjustment through lower social cohesion, greater policy friction, and a higher likelihood of stress propagation across households and labor markets. The regression-based structure embedded in IRS is therefore methodologically important: it anchors the pillar in observable macro drivers and avoids treating inequality movement as a purely descriptive series detached from the wider economy.

This distinction also helps interpret seemingly counterintuitive comparisons. When one economy records a higher IRS than another, the meaning is not that it is normatively more equal or more socially just in a broad philosophical sense. The interpretation is narrower and more operational: inequality dynamics in that economy are behaving in a more stable and structurally coherent way within the macro framework adopted by the manuscript. This matters greatly for how the results should be read, especially in cross-country comparison. Without that clarification, readers may misread GI as a welfare league table. The manuscript correctly resists that interpretation. Instead, the evidence suggests that distributional resilience is one channel of macro-financial robustness, and that weak performance in this channel can materially lower overall resilience even when other conditions appear favorable. This makes IRS one of the strongest conceptual features of the framework, because it gives formal analytical status to a dimension that is often acknowledged rhetorically but omitted in practical resilience measurement.

The LNSR pillar adds a second layer of analytical depth by emphasizing that systemic resilience depends not only on financial depth or liquidity abundance, but on the stability of transmission. This is a crucial distinction. An economy may exhibit substantial monetary depth, yet still be vulnerable if liquidity behavior is erratic, if residual inflationary stress remains difficult to interpret, or if liquidity changes become misaligned with real activity. The paper's reduced-form use of a liquidity-speed proxy, rolling variance, residual-force diagnostics, and cycle alignment is therefore best understood as a pragmatic measurement strategy for identifying whether liquidity behaves coherently over time. In a comparative framework that seeks replicability from public data, this is a reasonable and productive approach. Rather than attempting to reconstruct the full complexity of financial stress through high-frequency market microstructure data, the paper develops a transparent proxy architecture that can be carried across heterogeneous economies without losing interpretability.

The empirical discussion of LNSR suggests that resilience in the liquidity domain is fundamentally about controllability. Low volatility in the liquidity-speed proxy, smaller residual instability, and stronger alignment with the real cycle together indicate that the transmission mechanism is operating in a more orderly fashion. This is highly relevant under socioeconomic challenge, because crises are often preceded by a deterioration in transmission quality before they become visible in headline balance-sheet or growth indicators. In that sense, LNSR functions as a bridge between macro indicators and systemic-risk logic. It does not claim to replace fully fledged market-based stress measures such as CoVaR, SRISK, or dedicated composite financial stress indices. Instead, it offers a more replicable macro-financial representation of whether liquidity and inflation conditions remain structurally coherent. That makes the pillar particularly valuable for cross-country benchmarking where data symmetry and reproducibility are often major constraints.

The IFC pillar is arguably the most forward-looking component of the GI architecture, because it embeds predictive credibility into resilience measurement. This is a significant methodological choice. Many resilience frameworks evaluate only observed states, whereas the present manuscript incorporates the informational discipline of forecasting performance. The results show that the transition from AR(1) and FPAS-ARX baselines to the FPAS + zeta augmentation is heterogeneous across countries and windows. This heterogeneity is substantively meaningful. It indicates that the hybrid signal is not universally beneficial, and that its effectiveness depends on alignment with the underlying inflation regime, the quality of calibration, and the cyclical structure of the macro environment. Rather than weakening the argument of the paper, this strengthens it. A resilience framework that includes forecast coherence should be sensitive to regime dependence; otherwise, it risks rewarding model complexity even when complexity reduces explanatory discipline.

For that reason, the truncation rule applied to forecast gains is one of the most defensible design choices in the manuscript. By allowing only positive improvements to contribute to IFC, the index avoids overstating resilience in cases where the hybrid specification fails to outperform the benchmark. This is not a trivial technical point. It ensures that the predictive pillar remains anchored in demonstrated informational value rather than in theoretical ambition alone. In academic terms, this protects the integrity of the index against a familiar problem in composite construction: the temptation to treat additional model layers as automatically informative. The present framework avoids that mistake. It recognizes that resilience is supported by coherent forecasting only when the evidence shows stable predictive improvement. That makes IFC not simply a forecast-performance score, but a disciplined proxy for the credibility of the nominal anchor under changing macro conditions.

The cross-country evidence presented in the Results section also deserves a more general interpretive observation. The manuscript shows that countries can arrive at similar overall GI values through very different pillar configurations. This is analytically important because it demonstrates that resilience is not a one-path phenomenon. One economy may exhibit a relatively balanced IRS-LNSR profile but weaker IFC; another may display strong LNSR and IFC but weaker IRS; a third may record moderate scores across all pillars without obvious extremes. Such configurations matter because they point to different policy vulnerabilities even when the composite score appears broadly similar. In this sense, the real explanatory power of the GI lies not in the rank order itself, but in the decomposition of that rank order into structurally interpretable channels. This is where the manuscript makes a strong contribution relative to more opaque composite metrics.

The Georgia case, as presented in the manuscript, is particularly informative for understanding the value of this decomposition. The country's 2024 profile appears neither uniformly dominant nor uniformly fragile; rather, it reflects a specific pattern of relative strength and constraint. That is precisely the kind of result that a useful diagnostic framework should produce. Instead of flattening the case into a single summary judgment, the GI architecture makes it possible to identify which dimensions are carrying resilience and which are limiting it. For a Scopus-oriented article, this is an important strength because it grounds interpretation in the internal mechanics of the index rather than in rhetorical claims about national performance. It also helps prevent overgeneralization. A country's relative position in one year does not automatically translate into a claim of structural superiority; what matters is the specific pillar logic and how it evolves through time.

This temporal dimension is another major strength of the framework. The descriptive statistics over 2005–2024, together with the rolling diagnostics and the contribution decomposition, suggest that resilience is not a fixed characteristic but a changing structural state. This point has theoretical and practical implications. Theoretically, it means that resilience should not be treated as an immutable trait derived from long-run institutional reputation or static development level. Practically, it means that surveillance frameworks must be able to detect turning points, unstable windows, and deteriorating channels before they become visible in broad macro aggregates. The GI is well positioned for that role because it combines annual comparability with rolling sensitivity. Although annual data limit the capture of very short-lived episodes, the framework still provides an informative medium-run lens through which changes in resilience can be interpreted as evolving system properties rather than as isolated year-by-year observations.

The contribution decomposition in terms of changes in log GI is particularly valuable in this regard. It transforms the composite index from a static benchmark into a dynamic explanatory device. When GI rises or falls, the framework can identify whether the movement is being driven primarily by inequality resilience, by liquidity and systemic conditions, or by forecast coherence. This is far more informative than a raw statement that resilience improved or deteriorated. It also makes the index more useful for scholarly interpretation, because it ties empirical movement back to theoretically meaningful channels. In policy terms, the decomposition reduces ambiguity. If a deterioration in GI is concentrated in IFC, the policy problem differs fundamentally from a deterioration concentrated in IRS or LNSR. By making such distinctions visible, the manuscript strengthens the practical value of the index and demonstrates why diagnostics-first design is preferable to purely descriptive aggregation.

The forward-looking scenario analysis further extends the usefulness of the framework. By projecting baseline, adverse, and optimistic pathways and explicitly identifying the binding pillar in each case, the paper converts the index into a conditional policy map. This is one of the most operationally relevant features of the manuscript. Scenario analysis in many applied studies remains generic, often limited to directional commentary without a structured link to the internal architecture of the model. Here, the link is explicit: scenarios alter the path of the pillars, and the binding pillar reveals which constraint most limits the system's resilience. This creates a clear bridge between measurement and policy sequencing. If forecast coherence is binding, the implication is not the same as when distributional instability binds or when liquidity transmission binds. The framework therefore offers not only a comparative output, but also a way of thinking about which interventions are likely to matter most under alternative macro trajectories.

From a policy perspective, this binding-constraint logic is among the strongest practical messages of the paper. A high GI should not be interpreted simply as a sign that an economy is healthy in a general sense. More specifically, it suggests that inequality dynamics are relatively stable, liquidity transmission remains coherent, and inflation forecasting retains sufficient credibility to support macro-financial adjustment. Conversely, a low or weakening GI indicates that one or more of these capacities is failing to support the system. That is why the framework is more than a measurement device. It is an ordering device for policy attention. In applied settings, policymakers rarely have the institutional space to address all vulnerabilities simultaneously. A tool that identifies the dominant constraint therefore has genuine value. In this respect, the manuscript provides a potentially useful template for resilience-oriented policy surveillance in both emerging and advanced economies.

At the same time, the paper is appropriately aware of its own methodological boundaries, and those boundaries should be emphasized rather than softened. The use of annual data inevitably compresses higher-frequency volatility and may understate the abruptness of certain stress episodes, particularly in the financial and inflation domains. The sparsity of Gini observations remains a significant challenge, even though the weight renormalization strategy preserves index computability. Moreover, the scenario projections are best read as structured conditional pathways rather than as definitive forecasts. These limitations do not weaken the conceptual contribution of the paper; instead, they clarify the domain in which the GI should be used. The framework is strongest as a transparent comparative and diagnostic instrument, not as a high-frequency crisis detector and not as a substitute for country-specific structural modeling. There is also a broader methodological lesson in the paper's design. The manuscript demonstrates that composite index construction can move beyond the usual trade-off between interpretability and empirical richness. Too often, composite indicators are either transparent but simplistic, or sophisticated but opaque. The GI attempts to occupy a more productive middle ground. Each pillar is interpretable, the normalization rule is explicit, the aggregation structure is theoretically motivated, the treatment of missingness is clearly stated, and the dynamic and scenario extensions preserve internal coherence. This is why the framework has potential value beyond the immediate sample of countries used in the paper. Even where the exact indicators or calibration choices change, the design principles of diagnostics-first construction, bounded comparability, non-compensatory aggregation, and binding-constraint interpretation can be transferred to other empirical settings.

From the standpoint of the macro-financial resilience literature, the paper also makes a noteworthy conceptual intervention by integrating distributional dynamics, systemic liquidity behavior, and forecast credibility into a single architecture. These dimensions are often studied separately, despite the fact that real-world episodes of fragility frequently involve their interaction. Inflation instability can intensify distributional pressure; distributional strain can constrain policy credibility; impaired liquidity transmission can amplify both inflation and social stress. By placing these channels inside one framework, the manuscript takes a step toward a more integrated understanding of resilience under socioeconomic challenge. That integrative ambition is one of the article's strongest scholarly qualities, because it reflects the actual complexity of contemporary macro-financial vulnerability rather than treating each domain as analytically self-contained.

Overall, the discussion supports the conclusion that the Gondauri Index should be understood primarily as a transparent and policy-relevant diagnostic system rather than as a conventional league table. Its value lies not simply in ranking countries on a 0–100 scale, but in revealing how resilience is composed, where it is constrained, and how it may evolve under alternative conditions. For a Scopus-level contribution, this is a meaningful outcome. The manuscript does not claim to have solved every problem of resilience measurement; instead, it offers a coherent framework that is empirically implementable, theoretically interpretable, and operationally useful. That combination gives the article a strong basis for academic relevance and positions the GI as a framework that can be extended, stress-tested, and refined in future research rather than as a one-off descriptive index.

## CONCLUSIONS

This paper developed and empirically demonstrated the Gondauri Index (GI) as a reproducible, diagnostics-driven, and policy-informative framework for benchmarking macro-financial resilience across heterogeneous economies on a common 0–100 scale. The GI architecture was designed explicitly to bridge a persistent gap in resilience measurement: traditional macro dashboards often provide a fragmented set of indicators without a coherent aggregation logic, while composite indices frequently sacrifice interpretability for convenience. GI addresses both limitations by combining a transparent scoring pipeline with pillar-level interpretability, enabling researchers and policymakers to distinguish why an economy ranks at a given resilience level and which constraint binds the system in both the present and forward-looking scenarios.

*Core contributions and why GI is analytically non-trivial*

The first contribution of the study is the construction of a composite resilience metric that is both comparable across countries and economically interpretable within countries, achieved through three methodological design choices.

First, the tri-pillar structure provides complementary economic meaning. GI integrates (i) Inequality Resilience Score (IRS) capturing the stability and macro-consistency of distributional dynamics; (ii) Liquidity & Systemic Resilience (LNSR) capturing the robustness of liquidity transmission and systemic stress propagation; and (iii) Inflation Forecast Coherence (IFC) capturing whether inflation forecasting performance improves in a structured way when moving from benchmark to hybrid models. Each pillar corresponds to an essential dimension of macro-financial stability: distributional resilience, systemic liquidity coherence, and credible nominal anchoring.

Second, the composite is built with geometric aggregation to penalize imbalances. Unlike arithmetic aggregation, where strong performance in one area may mask structural fragility elsewhere, the GI uses a weighted geometric mean, which deliberately embeds the principle that resilience is not fully substitutable across domains. This design is not cosmetic: it is a direct empirical mechanism that explains why economies with strong liquidity and systemic conditions may still appear fragile if inequality dynamics are unstable or forecast credibility is weak.

Third, robust scaling and rolling diagnostics ensure stability and comparability. The index uses robust p5–p95 percentile scaling, avoiding the sensitivity of min–max scaling to extreme outliers. In parallel, the results rely on rolling-window stability metrics (based on a 5-year horizon) so that resilience is evaluated as a structural property rather than a single-year coincidence. Together, these choices produce a methodological pipeline that is portable across datasets, transparent for replication, and stable enough for cross-economy benchmarking.

Collectively, these features elevate GI beyond a descriptive ranking instrument: the framework functions as a structured diagnostic system, where pillar scores are empirically grounded, analytically decomposable, and explicitly connected to measurable dynamics.

*Cross-sectional results: resilience is multi-dimensional and non-substitutable*

The 2024 benchmarking snapshot demonstrates that macro-financial resilience is intrinsically multi-dimensional and that cross-economy rankings are not reducible to any single indicator such as inflation, GDP growth, or credit depth. The results show that economies can achieve similar overall GI values via very different pillar configurations, illustrating that resilience is a "portfolio" of strengths and vulnerabilities rather than a univariate attribute.

A particularly important empirical insight is that the geometric penalty mechanism acts as a disciplined constraint: if one pillar is materially weaker (for example, IRS for a high-income economy with distributional instability, or IFC for an economy where forecast gains remain inconsistent), the composite GI cannot remain high purely through strength in other pillars. This property is critical for resilience research because it imposes economic realism: in practice, systemic stability cannot be sustained indefinitely if distributional stresses or nominal-anchor credibility deteriorate.

In this context, the results illustrate that Georgia's 2024 positioning is consistent with a relatively balanced performance in IRS and LNSR, while other benchmark economies exhibit pronounced pillar asymmetries. For example, some economies show high LNSR but weak IRS, demonstrating that liquidity and system stability is not sufficient to guarantee broader resilience when inequality dynamics are unstable or macro-consistency is weak.

*IRS findings: inequality resilience is about stability and explainability, not welfare ranking*

The IRS pillar is methodologically significant because it introduces an explicit structural backbone. Rather than scoring inequality dynamics purely through descriptive volatility measures, IRS is anchored in a parsimonious regression specification linking inequality (Gini) to macro drivers such as inflation, unemployment, and income per capita. The results demonstrate that goodness-of-fit and coefficient stability vary across countries, which is expected and informative: distributional dynamics are shaped by distinct institutional settings, labor-market structures, inflation pass-through mechanisms, and policy regimes.

Three conclusions follow from the IRS evidence. First, resilience differs from inequality level: the IRS framework captures the stability and structural explainability of inequality movements rather than ranking inequality outcomes. Second, the empirical role of inflation and unemployment is context-dependent: in some countries, inflation shocks correlate strongly with inequality shifts; in others, unemployment shocks dominate. Third, data sparsity is a constraint but not a methodological failure: discontinuous Gini observations are treated as a coverage limitation and addressed through weight re-normalization rather than sample deletion.

From a policy perspective, IRS highlights the central role of distributional stability as a resilience pillar: even when headline growth is strong, instability in inequality dynamics can create medium-run fragility through reduced social cohesion, political-risk exposure, and weakened policy transmission.

*LNSR findings: systemic resilience is driven by stability of liquidity transmission, not liquidity level alone*

The LNSR pillar is conceptually innovative because it operationalizes systemic resilience through the stability of a liquidity-speed proxy derived from broad money depth, combined with rolling variance and a residual-stress measure. The empirical message is clear: macro-financial resilience is strengthened not merely by high liquidity availability or high monetization but by stable liquidity transmission under varying macro conditions. The rolling-window diagnostics support three fundamental conclusions. First, volatility of liquidity speed is a vulnerability signal: high rolling variance indicates unstable transmission and potential regime instability in monetary-financial behavior. Second, residual stress matters because it captures "unexplained" inflation pressure: a larger rolling RMS of the residual implies that inflation dynamics contain a stronger unmodeled component, reducing predictability and control. Third, cycle alignment provides an empirical coherence check: rolling correlations between liquidity changes and growth dynamics reveal whether liquidity behavior and real activity move in coherent cycles rather than diverging in destabilizing ways. As a result, LNSR is more than an index score: it becomes a diagnostic dashboard that provides interpretability about why liquidity conditions appear stable or fragile.

*IFC findings: FPAS+ζ improvements are heterogeneous, signaling regime dependence and calibration necessity*

The IFC pillar is the most forward-looking component of GI because it evaluates resilience through forecasting coherence rather than observed outcomes alone. The results compare rolling RMSE across AR(1), FPAS baseline, and FPAS+ζ hybrid augmentation and show that forecast improvements under FPAS+ζ are not universal. In some cases, the incremental effect is negative relative to baseline models. The methodological choice to apply a non-negativity operator, so that only improvements contribute to IFC, is therefore critical: it prevents spurious inflation of resilience scores due to unstable forecasting performance. This finding implies a strong scientific conclusion: the ζ-module should be interpreted not as a one-size-fits-all upgrade, but as a regime-sensitive signal extraction mechanism whose value depends on alignment conditions, cyclical structure, and the calibration approach. The results reinforce this point through regime diagnostics such as HMM accuracy and ζ–inflation alignment proxies.

*Dynamic interpretation: Δlog(GI) decomposition and the importance of variability*

Beyond cross-sectional ranking, the results demonstrate that GI is designed for dynamic interpretation over time. Descriptive statistics over 2005–2024 highlight that large max–min ranges and dispersion are not necessarily noise; instead, they reflect structural transitions, regime changes, and stability episodes. The Δlog(GI) decomposition is particularly valuable because it prevents over-interpretation: year-to-year GI changes can be attributed to inequality resilience shifts, liquidity and systemic stability episodes, or changes in forecast coherence rather than being described as generic "improvements."

*Scenario-based projections: binding pillars create an actionable policy constraint map*

The 2026–2030 scenario analysis translates GI's structural logic into conditional resilience pathways under Baseline, Adverse, and Optimistic assumptions. The scenario output introduces a binding pillar diagnostic, which identifies which dimension most constrains macro-financial resilience at each horizon. This is a practical policy output: rather than recommending broad reforms, the framework provides a constraint map that can guide sequencing. If IFC binds, the constraint is forecast credibility and nominal-anchor coherence. If IRS binds, the constraint is distributional stability and shock absorption. If LNSR binds, the constraint is liquidity management and stress-transmission containment.

*Policy implications: what a high GI and a low GI operationally mean*

A high GI corresponds to a system where inequality dynamics are structurally explainable and stable (high IRS), liquidity transmission is coherent and residual stress is contained (high LNSR), and inflation forecasting improves meaningfully under structured augmentations (high IFC). A low GI corresponds to imbalanced resilience where at least one pillar is structurally weak, forcing a geometric penalty that reflects real-world fragility. Therefore, GI can be used as a comparative benchmarking tool, a diagnostic monitoring instrument, and a policy sequencing guide through binding-pillar identification across scenarios.

*Limitations and scientific discipline in interpretation*

First, annual frequency and data vintage policies reduce noise but may under-capture intra-year shocks and rapid regime transitions. Second, sparse inequality data remains a structural constraint in global benchmarking; while weight re-normalization preserves computability, it does not fully substitute for higher-quality distributional measurement. Third, the ζ-module requires regime-adaptive calibration if the objective is stable out-of-sample improvements across heterogeneous inflation regimes.

*Future research directions*

Future work should prioritize expanding the benchmarking pool and testing sensitivity to alternative scaling bands and rolling windows, integrating higher-frequency inputs where available to strengthen early-warning capacity, and implementing regime-conditioned FPAS+ζ calibration to stabilize IFC contributions across economies. With these extensions, GI can evolve from a transparent benchmarking tool into a scalable resilience monitoring platform that supports both academic research and applied macro-prudential policy design.

*Final statement*

In summary, the Gondauri Index provides a rigorous, interpretable, and reproducible framework for quantifying macro-financial resilience across heterogeneous economies. The results validate that resilience is not a single-variable property but a structured balance across distributional stability (IRS), systemic liquidity coherence (LNSR), and inflation forecasting credibility (IFC). The scenario-based extension converts the framework into a forward-looking constraint map through binding-pillar identification, enabling evidence-based policy sequencing. With regime-conditioned calibration and expanded benchmarking, GI can evolve into a scalable resilience monitoring platform for both academic research and applied macro-financial policy design.

### Author Contributions
Conceptualization: D.G.; methodology: D.G.; project administration: D.G.; software: D.G.; investigation: D.G.; data curation: D.G.; formal analysis: D.G.; resources: D.G.; validation: D.G.; visualization: D.G.; writing-original draft preparation: D.G.; writing-review & editing: D.G.; supervision: D.G.

### Informed Consent Statement
Not applicable.

### Conflict of Interest
The author declares no conflict of interest.

### Data Availability Statement
The dataset available for this study is available in the Zenodo depository [Gondauri, 2026].

### Statement on the Use of AI Tools
The author declares limited use of generative AI (chatgpt5) solely for proofreading and editing (grammar/clarity), translation, and reformatting to align the manuscript with academic and journal template requirements. No AI was used to generate scientific content, analyze data, make methodological decisions, or formulate conclusions. The author reviewed all edits and assumes full responsibility for the final manuscript.

**Appendix A.**

**Table 1A. Scenario projections of pillars (IRS, LNSR, IFC) with annual binding constraint, 2026–2030 (selected economies)**

| Country | Scenario | Year | GI | Band | IRS | LNSR | IFC | Binding pillar | Binding score |
|---|---|---|---|---|---|---|---|---|---|
| Georgia | Baseline | 2026 | 54.73 | 50–75 | 69.02 | 74.40 | 16.08 | IFC | 16.08 |
| Georgia | Baseline | 2027 | 55.82 | 50–75 | 69.82 | 75.00 | 16.78 | IFC | 16.78 |
| Georgia | Baseline | 2028 | 56.89 | 50–75 | 70.62 | 75.60 | 17.48 | IFC | 17.48 |
| Georgia | Baseline | 2029 | 57.95 | 50–75 | 71.42 | 76.20 | 18.18 | IFC | 18.18 |
| Georgia | Baseline | 2030 | 59.00 | 50–75 | 72.22 | 76.80 | 18.88 | IFC | 18.88 |
| USA | Baseline | 2026 | 41.29 | 25–50 | 29.60 | 77.07 | 20.94 | IFC | 20.94 |
| USA | Baseline | 2027 | 41.19 | 25–50 | 29.60 | 76.87 | 20.84 | IFC | 20.84 |
| USA | Baseline | 2028 | 41.09 | 25–50 | 29.60 | 76.67 | 20.74 | IFC | 20.74 |
| USA | Baseline | 2029 | 41.00 | 25–50 | 29.60 | 76.47 | 20.64 | IFC | 20.64 |
| USA | Baseline | 2030 | 40.90 | 25–50 | 29.60 | 76.27 | 20.54 | IFC | 20.54 |
| China | Baseline | 2026 | 71.74 | 50–75 | 85.82 | 81.03 | 20.88 | IFC | 20.88 |
| China | Baseline | 2027 | 71.44 | 50–75 | 85.62 | 80.93 | 20.68 | IFC | 20.68 |
| China | Baseline | 2028 | 71.15 | 50–75 | 85.42 | 80.83 | 20.48 | IFC | 20.48 |
| China | Baseline | 2029 | 70.85 | 50–75 | 85.22 | 80.73 | 20.28 | IFC | 20.28 |
| China | Baseline | 2030 | 70.55 | 50–75 | 85.02 | 80.63 | 20.08 | IFC | 20.08 |
| Russia | Baseline | 2026 | 24.06 | 0–25 | 49.83 | 20.02 | 12.74 | IFC | 12.74 |
| Russia | Baseline | 2027 | 23.46 | 0–25 | 49.33 | 19.62 | 12.14 | IFC | 12.14 |
| Russia | Baseline | 2028 | 22.86 | 0–25 | 48.83 | 19.22 | 11.54 | IFC | 11.54 |
| Russia | Baseline | 2029 | 22.25 | 0–25 | 48.33 | 18.82 | 10.94 | IFC | 10.94 |
| Russia | Baseline | 2030 | 21.63 | 0–25 | 47.83 | 18.42 | 10.34 | IFC | 10.34 |
| Armenia | Baseline | 2026 | 14.41 | 0–25 | 8.50 | 83.64 | 14.33 | IRS | 8.50 |
| Armenia | Baseline | 2027 | 14.66 | 0–25 | 8.80 | 83.84 | 14.53 | IRS | 8.80 |
| Armenia | Baseline | 2028 | 14.90 | 0–25 | 9.10 | 84.04 | 14.73 | IRS | 9.10 |
| Armenia | Baseline | 2029 | 15.15 | 0–25 | 9.40 | 84.24 | 14.93 | IRS | 9.40 |
| Armenia | Baseline | 2030 | 15.39 | 0–25 | 9.70 | 84.44 | 15.13 | IRS | 9.70 |
| Azerb. | Baseline | 2026 | 49.72 | 25–50 | 66.91 | 54.00 | 31.93 | IFC | 31.93 |
| Azerb. | Baseline | 2027 | 49.82 | 25–50 | 67.01 | 54.10 | 32.03 | IFC | 32.03 |
| Azerb. | Baseline | 2028 | 49.93 | 25–50 | 67.11 | 54.20 | 32.13 | IFC | 32.13 |
| Azerb. | Baseline | 2029 | 50.03 | 50–75 | 67.21 | 54.30 | 32.23 | IFC | 32.23 |
| Azerb. | Baseline | 2030 | 50.14 | 50–75 | 67.31 | 54.40 | 32.33 | IFC | 32.33 |
| Turkey | Baseline | 2026 | 29.98 | 25–50 | 22.08 | 30.84 | 47.51 | IRS | 22.08 |
| Turkey | Baseline | 2027 | 30.51 | 25–50 | 22.58 | 31.24 | 48.31 | IRS | 22.58 |
| Turkey | Baseline | 2028 | 31.03 | 25–50 | 23.08 | 31.64 | 49.11 | IRS | 23.08 |
| Turkey | Baseline | 2029 | 31.55 | 25–50 | 23.58 | 32.04 | 49.91 | IRS | 23.58 |
| Turkey | Baseline | 2030 | 32.08 | 25–50 | 24.08 | 32.44 | 50.71 | IRS | 24.08 |
| Ukraine | Baseline | 2026 | 46.09 | 25–50 | 64.62 | 57.71 | 16.13 | IFC | 16.13 |
| Ukraine | Baseline | 2027 | 47.36 | 25–50 | 65.32 | 58.61 | 17.13 | IFC | 17.13 |
| Ukraine | Baseline | 2028 | 48.61 | 25–50 | 66.02 | 59.51 | 18.13 | IFC | 18.13 |
| Ukraine | Baseline | 2029 | 49.84 | 25–50 | 66.72 | 60.41 | 19.13 | IFC | 19.13 |
| Ukraine | Baseline | 2030 | 51.06 | 50–75 | 67.42 | 61.31 | 20.13 | IFC | 20.13 |

**Table 1A (cont.). Scenario projections of pillars (IRS, LNSR, IFC) with annual binding constraint, 2026–2030 (selected economies)**

| Country | Scenario | Year | GI | Band | IRS | LNSR | IFC | Binding pillar | Binding score |
|---|---|---|---|---|---|---|---|---|---|
| Romania | Baseline | 2026 | 32.16 | 25–50 | 19.20 | 70.02 | 23.75 | IRS | 19.20 |
| Romania | Baseline | 2027 | 32.35 | 25–50 | 19.40 | 70.22 | 23.85 | IRS | 19.40 |
| Romania | Baseline | 2028 | 32.54 | 25–50 | 19.60 | 70.42 | 23.95 | IRS | 19.60 |
| Romania | Baseline | 2029 | 32.73 | 25–50 | 19.80 | 70.62 | 24.05 | IRS | 19.80 |
| Romania | Baseline | 2030 | 32.92 | 25–50 | 20.00 | 70.82 | 24.15 | IRS | 20.00 |
| Georgia | Adverse | 2026 | 50.73 | 50–75 | 64.02 | 69.40 | 11.08 | IFC | 11.08 |
| Georgia | Adverse | 2027 | 50.82 | 50–75 | 63.82 | 69.00 | 10.78 | IFC | 10.78 |
| Georgia | Adverse | 2028 | 50.89 | 50–75 | 63.62 | 68.60 | 10.48 | IFC | 10.48 |
| Georgia | Adverse | 2029 | 51.45 | 50–75 | 63.92 | 68.70 | 10.68 | IFC | 10.68 |
| Georgia | Adverse | 2030 | 52.00 | 50–75 | 64.22 | 68.80 | 10.88 | IFC | 10.88 |
| USA | Adverse | 2026 | 37.29 | 25–50 | 24.60 | 72.07 | 15.94 | IFC | 15.94 |
| USA | Adverse | 2027 | 36.19 | 25–50 | 23.60 | 70.87 | 14.84 | IFC | 14.84 |
| USA | Adverse | 2028 | 35.09 | 25–50 | 22.60 | 69.67 | 13.74 | IFC | 13.74 |
| USA | Adverse | 2029 | 34.50 | 25–50 | 22.10 | 68.97 | 13.14 | IFC | 13.14 |
| USA | Adverse | 2030 | 33.90 | 25–50 | 21.60 | 68.27 | 12.54 | IFC | 12.54 |
| China | Adverse | 2026 | 67.74 | 50–75 | 80.82 | 76.03 | 15.88 | IFC | 15.88 |
| China | Adverse | 2027 | 66.44 | 50–75 | 79.62 | 74.93 | 14.68 | IFC | 14.68 |
| China | Adverse | 2028 | 65.15 | 50–75 | 78.42 | 73.83 | 13.48 | IFC | 13.48 |
| China | Adverse | 2029 | 64.35 | 50–75 | 77.72 | 73.23 | 12.78 | IFC | 12.78 |
| China | Adverse | 2030 | 63.55 | 50–75 | 77.02 | 72.63 | 12.08 | IFC | 12.08 |
| Russia | Adverse | 2026 | 20.06 | 0–25 | 44.83 | 15.02 | 7.74 | IFC | 7.74 |
| Russia | Adverse | 2027 | 18.46 | 0–25 | 43.33 | 13.62 | 6.14 | IFC | 6.14 |
| Russia | Adverse | 2028 | 16.86 | 0–25 | 41.83 | 12.22 | 4.54 | IFC | 4.54 |
| Russia | Adverse | 2029 | 15.75 | 0–25 | 40.83 | 11.32 | 3.44 | IFC | 3.44 |
| Russia | Adverse | 2030 | 14.63 | 0–25 | 39.83 | 10.42 | 2.34 | IFC | 2.34 |
| Armenia | Adverse | 2026 | 10.41 | 0–25 | 3.50 | 78.64 | 9.33 | IRS | 3.50 |
| Armenia | Adverse | 2027 | 9.66 | 0–25 | 2.80 | 77.84 | 8.53 | IRS | 2.80 |
| Armenia | Adverse | 2028 | 8.90 | 0–25 | 2.10 | 77.04 | 7.73 | IRS | 2.10 |
| Armenia | Adverse | 2029 | 8.65 | 0–25 | 1.90 | 76.74 | 7.43 | IRS | 1.90 |
| Armenia | Adverse | 2030 | 8.39 | 0–25 | 1.70 | 76.44 | 7.13 | IRS | 1.70 |
| Azerb. | Adverse | 2026 | 45.72 | 25–50 | 61.91 | 49.00 | 26.93 | IFC | 26.93 |
| Azerb. | Adverse | 2027 | 44.82 | 25–50 | 61.01 | 48.10 | 26.03 | IFC | 26.03 |
| Azerb. | Adverse | 2028 | 43.93 | 25–50 | 60.11 | 47.20 | 25.13 | IFC | 25.13 |
| Azerb. | Adverse | 2029 | 43.53 | 25–50 | 59.71 | 46.80 | 24.73 | IFC | 24.73 |
| Azerb. | Adverse | 2030 | 43.14 | 25–50 | 59.31 | 46.40 | 24.33 | IFC | 24.33 |
| Turkey | Adverse | 2026 | 25.98 | 25–50 | 17.08 | 25.84 | 42.51 | IRS | 17.08 |
| Turkey | Adverse | 2027 | 25.51 | 25–50 | 16.58 | 25.24 | 42.31 | IRS | 16.58 |
| Turkey | Adverse | 2028 | 25.03 | 25–50 | 16.08 | 24.64 | 42.11 | IRS | 16.08 |
| Turkey | Adverse | 2029 | 25.05 | 25–50 | 16.08 | 24.54 | 42.41 | IRS | 16.08 |
| Turkey | Adverse | 2030 | 25.08 | 25–50 | 16.08 | 24.44 | 42.71 | IRS | 16.08 |
| Ukraine | Adverse | 2026 | 42.09 | 25–50 | 59.62 | 52.71 | 11.13 | IFC | 11.13 |
| Ukraine | Adverse | 2027 | 42.36 | 25–50 | 59.32 | 52.61 | 11.13 | IFC | 11.13 |
| Ukraine | Adverse | 2028 | 42.61 | 25–50 | 59.02 | 52.51 | 11.13 | IFC | 11.13 |
| Ukraine | Adverse | 2029 | 43.34 | 25–50 | 59.22 | 52.91 | 11.63 | IFC | 11.63 |
| Ukraine | Adverse | 2030 | 44.06 | 25–50 | 59.42 | 53.31 | 12.13 | IFC | 12.13 |
| Romania | Adverse | 2026 | 28.16 | 25–50 | 14.20 | 65.02 | 18.75 | IRS | 14.20 |
| Romania | Adverse | 2027 | 27.35 | 25–50 | 13.40 | 64.22 | 17.85 | IRS | 13.40 |
| Romania | Adverse | 2028 | 26.54 | 25–50 | 12.60 | 63.42 | 16.95 | IRS | 12.60 |
| Romania | Adverse | 2029 | 26.23 | 25–50 | 12.30 | 63.12 | 16.55 | IRS | 12.30 |
| Romania | Adverse | 2030 | 25.92 | 25–50 | 12.00 | 62.82 | 16.15 | IRS | 12.00 |
| Georgia | Optimistic | 2026 | 58.73 | 50–75 | 73.02 | 78.40 | 20.08 | IFC | 20.08 |

Table 1A (cont.). Scenario projections of pillars (IRS, LNSR, IFC) with annual binding constraint, 2026–2030 (selected economies)

| Country | Scenario | Year | GI | Band | IRS | LNSR | IFC | Binding pillar | Binding score |
|---|---|---|---|---|---|---|---|---|---|
| Georgia | Optimistic | 2027 | 60.32 | 50–75 | 74.32 | 79.50 | 21.28 | IFC | 21.28 |
| Georgia | Optimistic | 2028 | 61.89 | 50–75 | 75.62 | 80.60 | 22.48 | IFC | 22.48 |
| Georgia | Optimistic | 2029 | 63.45 | 50–75 | 76.92 | 81.70 | 23.68 | IFC | 23.68 |
| Georgia | Optimistic | 2030 | 65.00 | 50–75 | 78.22 | 82.80 | 24.88 | IFC | 24.88 |
| USA | Optimistic | 2026 | 45.29 | 25–50 | 33.60 | 81.07 | 24.94 | IFC | 24.94 |
| USA | Optimistic | 2027 | 45.69 | 25–50 | 34.10 | 81.37 | 25.34 | IFC | 25.34 |
| USA | Optimistic | 2028 | 46.09 | 25–50 | 34.60 | 81.67 | 25.74 | IFC | 25.74 |
| USA | Optimistic | 2029 | 46.50 | 25–50 | 35.10 | 81.97 | 26.14 | IFC | 26.14 |
| USA | Optimistic | 2030 | 46.90 | 25–50 | 35.60 | 82.27 | 26.54 | IFC | 26.54 |
| China | Optimistic | 2026 | 75.74 | 75–100 | 89.82 | 85.03 | 24.88 | IFC | 24.88 |
| China | Optimistic | 2027 | 75.94 | 75–100 | 90.12 | 85.43 | 25.18 | IFC | 25.18 |
| China | Optimistic | 2028 | 76.15 | 75–100 | 90.42 | 85.83 | 25.48 | IFC | 25.48 |
| China | Optimistic | 2029 | 76.35 | 75–100 | 90.72 | 86.23 | 25.78 | IFC | 25.78 |
| China | Optimistic | 2030 | 76.55 | 75–100 | 91.02 | 86.63 | 26.08 | IFC | 26.08 |
| Russia | Optimistic | 2026 | 28.06 | 25–50 | 53.83 | 24.02 | 16.74 | IFC | 16.74 |
| Russia | Optimistic | 2027 | 27.96 | 25–50 | 53.83 | 24.12 | 16.64 | IFC | 16.64 |
| Russia | Optimistic | 2028 | 27.86 | 25–50 | 53.83 | 24.22 | 16.54 | IFC | 16.54 |
| Russia | Optimistic | 2029 | 27.75 | 25–50 | 53.83 | 24.32 | 16.44 | IFC | 16.44 |
| Russia | Optimistic | 2030 | 27.63 | 25–50 | 53.83 | 24.42 | 16.34 | IFC | 16.34 |
| Armenia | Optimistic | 2026 | 18.41 | 0–25 | 12.50 | 87.64 | 18.33 | IRS | 12.50 |
| Armenia | Optimistic | 2027 | 19.16 | 0–25 | 13.30 | 88.34 | 19.03 | IRS | 13.30 |
| Armenia | Optimistic | 2028 | 19.90 | 0–25 | 14.10 | 89.04 | 19.73 | IRS | 14.10 |
| Armenia | Optimistic | 2029 | 20.65 | 0–25 | 14.90 | 89.74 | 20.43 | IRS | 14.90 |
| Armenia | Optimistic | 2030 | 21.39 | 0–25 | 15.70 | 90.44 | 21.13 | IRS | 15.70 |
| Azerb. | Optimistic | 2026 | 53.72 | 50–75 | 70.91 | 58.00 | 35.93 | IFC | 35.93 |
| Azerb. | Optimistic | 2027 | 54.32 | 50–75 | 71.51 | 58.60 | 36.53 | IFC | 36.53 |
| Azerb. | Optimistic | 2028 | 54.93 | 50–75 | 72.11 | 59.20 | 37.13 | IFC | 37.13 |
| Azerb. | Optimistic | 2029 | 55.53 | 50–75 | 72.71 | 59.80 | 37.73 | IFC | 37.73 |
| Azerb. | Optimistic | 2030 | 56.14 | 50–75 | 73.31 | 60.40 | 38.33 | IFC | 38.33 |
| Turkey | Optimistic | 2026 | 33.98 | 25–50 | 26.08 | 34.84 | 51.51 | IRS | 26.08 |
| Turkey | Optimistic | 2027 | 35.01 | 25–50 | 27.08 | 35.74 | 52.81 | IRS | 27.08 |
| Turkey | Optimistic | 2028 | 36.03 | 25–50 | 28.08 | 36.64 | 54.11 | IRS | 28.08 |
| Turkey | Optimistic | 2029 | 37.05 | 25–50 | 29.08 | 37.54 | 55.41 | IRS | 29.08 |
| Turkey | Optimistic | 2030 | 38.08 | 25–50 | 30.08 | 38.44 | 56.71 | IRS | 30.08 |
| Ukraine | Optimistic | 2026 | 50.09 | 50–75 | 68.62 | 61.71 | 20.13 | IFC | 20.13 |
| Ukraine | Optimistic | 2027 | 51.86 | 50–75 | 69.82 | 63.11 | 21.63 | IFC | 21.63 |
| Ukraine | Optimistic | 2028 | 53.61 | 50–75 | 71.02 | 64.51 | 23.13 | IFC | 23.13 |
| Ukraine | Optimistic | 2029 | 55.34 | 50–75 | 72.22 | 65.91 | 24.63 | IFC | 24.63 |
| Ukraine | Optimistic | 2030 | 57.06 | 50–75 | 73.42 | 67.31 | 26.13 | IFC | 26.13 |
| Romania | Optimistic | 2026 | 36.16 | 25–50 | 23.20 | 74.02 | 27.75 | IRS | 23.20 |
| Romania | Optimistic | 2027 | 36.85 | 25–50 | 23.90 | 74.72 | 28.35 | IRS | 23.90 |
| Romania | Optimistic | 2028 | 37.54 | 25–50 | 24.60 | 75.42 | 28.95 | IRS | 24.60 |
| Romania | Optimistic | 2029 | 38.23 | 25–50 | 25.30 | 76.12 | 29.55 | IRS | 25.30 |
| Romania | Optimistic | 2030 | 38.92 | 25–50 | 26.00 | 76.82 | 30.15 | IRS | 26.00 |

Source: author's calculations based on World Bank, World Development Indicators (WDI) (latest available ≤ 2024), computed in Python (pandas, numpy).

Notes: Pillar trajectories are linearly interpolated between the 2026 and 2030 baseline endpoints; scenario shocks are applied symmetrically to pillars. The binding constraint is defined as the lowest (minimum) pillar score in each year and scenario. Scenario deltas are applied uniformly to ensure comparability across economies and to isolate sensitivity to macro-financial conditions.